\documentclass[
  aps,
  pra,
  reprint,           
  onecolumn,         
  superscriptaddress,
  amsmath,amssymb,
  nofootinbib,
  longbibliography,
  floatfix
]{revtex4-2}

\usepackage{graphicx}
\usepackage{amsfonts}
\usepackage{mathrsfs}
\usepackage{xcolor}
\usepackage{booktabs}
\usepackage{multirow}
\usepackage{tikz}
\usepackage{adjustbox}
\usepackage[normalem]{ulem}
\usepackage{etoolbox}
\usepackage{hyperref}

\providecommand{\ket}{}
\renewcommand{\ket}[1]{\left\vert #1 \right\rangle}

\begin{document}

\title{Ancilla-assisted nondestructive discrimination of distributed GHZ-class states}

\author{Abdul Q. Batin}
\email{aqbatin@gmail.com}
\affiliation{Centre for Quantum Science and Technology, Siksha O Anusandhan Deemed to be University, Bhubaneswar 751030, Odisha, India}

\author{Suman Chand}
\email{sumanchand@soa.ac.in}
\affiliation{Centre for Quantum Science and Technology, Siksha O Anusandhan Deemed to be University, Bhubaneswar 751030, Odisha, India}

\author{Rajiuddin Sk}
\email{skrajiuddin@gmail.com}
\affiliation{Centre for Quantum Science and Technology, Siksha O Anusandhan Deemed to be University, Bhubaneswar 751030, Odisha, India}

\author{Prasanta K. Panigrahi}
\email{director.cqst@soa.ac.in}
\affiliation{Centre for Quantum Science and Technology, Siksha O Anusandhan Deemed to be University, Bhubaneswar 751030, Odisha, India}

\date{\today}

\begin{abstract}
Nondestructive quantum state discrimination is a fundamental primitive in distributed quantum information processing, where shared multipartite entangled resources need to be identified without being consumed. In this work, we present a scalable ancilla-assisted protocol for the strict nondestructive discrimination of $n$-qubit GHZ-class states distributed among distant parties. By employing multipartite GHZ ancillary states and local unitary interactions, we show that the computational-pattern and relative-phase information of the system GHZ state can be coherently mapped onto two ancillary registers while leaving the system state unchanged. We explicitly derive the discrimination rules for three- and four-qubit GHZ-class states and develop a systematic extension to arbitrary $n$. We further investigate the discrimination protocol in the presence of depolarizing and amplitude-damping noise in the ancillary system and obtain analytical expressions for the success probability and the corresponding critical noise thresholds. The protocol is also implemented using the IBM Qiskit platform to demonstrate its experimental feasibility and reproduce the expected ancillary measurement signatures. The proposed framework provides a scalable approach to the strict nondestructive discrimination of multipartite GHZ-class states and suggests a broader interpretation of the ancillary GHZ resources as probes of decoherence, opening a possible connection to quantum decoherence sensing in distributed quantum systems.
\end{abstract}


\maketitle

\section{Introduction}
Quantum state discrimination is a fundamental concept in quantum theory, serving as a foundation for several key principles including the no-cloning theorem \cite{wootters1982single,lindblad1999general,kalev2008no,wootters2009no} and quantum nonlocality \cite{popescu1992generic,popescu1994causality,popescu1994quantum,bennett1999quantum,cavalcanti2011quantum,bhattacharya2020nonlocality}. It plays a central role in quantum information processing, with applications ranging from quantum communication to cryptography. For instance, the security of quantum key distribution relies on the inherent impossibility of perfectly distinguishing non-orthogonal quantum states \cite{bennett2014quantum,pirandola2020advances,xu2020secure}. Measurements that discriminate maximally entangled states also play a central role in quantum information processing, enabling protocols such as quantum teleportation and entanglement swapping \cite{bouwmeester1997experimental,jin2015highly}. Extending such discrimination capabilities from bipartite to genuinely multipartite entangled states remains an important challenge for distributed quantum information processing because multipartite entanglement possesses qualitatively different correlation structures that cannot be reduced to independent pairwise correlations.

Conceptually, quantum state discrimination protocols can be broadly classified into destructive and nondestructive schemes \cite{chefles2000quantum, barnett2009quantum, bergou2010discrimination, bae2015quantum}. In the former scheme, the state is irreversibly collapsed during measurement, whereas the latter aims to identify the state while preserving it for further use. This distinction is particularly important in quantum networks \cite{kimble2008quantum,wehner2018quantum} and fault-tolerant quantum architectures, where entangled resources are costly to generate and must often be verified, reused, or distributed across multiple nodes without being consumed. In large-scale quantum networks, long-distance entanglement distribution has already been experimentally demonstrated, highlighting the importance of reliable methods for verifying shared entangled states without destroying them \cite{yin2017satellite}.

Discrimination of orthogonal entangled states is therefore of particular interest, and a variety of methods have been proposed for its realization \cite{gupta2007general,panigrahi2006circuits,wang2013nondestructive, satyajit2018nondestructive}. A widely used approach to nondestructive discrimination relies on unitary interactions between the target system and ancillary qubits, such that the relevant distinguishing information is coherently transferred to the ancilla for subsequent measurement, while leaving the system state intact \cite{jain2009secure,satyajit2018nondestructive,NondestructivePRA2025,lim2025trade}. Several works have demonstrated the feasibility of such ancilla-assisted nondestructive discrimination schemes, including experimental implementations and applications to stabilizer states \cite{gupta2024bell,samal2010non,sisodia2017experimental,ghosh2018automated,thatte2025quantum,welte2021nondestructive}.

In nondestructive discrimination schemes, an important distinction arises between protocols that rely solely on local ancillary systems and those that exploit entangled ancillary resources shared among distant parties. Protocols based on locally available ancillas are fundamentally constrained by local operations and classical communication (LOCC), which impose intrinsic limits on the discrimination of entangled states distributed across spatially separated nodes \cite{bennett1999quantum,walgate2000local,ghosh2001distinguishability}. For example, recent work has shown that while nondestructive Bell-state discrimination using local operations and classical communication (LOCC) alone cannot exceed the classical success probability of random guessing, these limitations can be overcome by employing preshared entangled ancillas to access nonlocal correlations \cite{NondestructivePRA2025}. In such settings, the ancillary entanglement effectively mediates delocalized quantum interactions between distant parties, enabling the extraction of global information about the shared state while preserving it \cite{paige2020quantum,horodecki2009quantum}.


Within this framework, complete nondestructive discrimination has been demonstrated for certain classes of maximally entangled states, most notably the Bell states \cite{gupta2007general,panigrahi2006circuits,wang2013nondestructive, satyajit2018nondestructive,NondestructivePRA2025}. This success is closely related to the stabilizer structure of Bell states, whose distinguishing information can be associated with parity and phase observables accessible through local interactions assisted by entangled ancillas \cite{gottesman1997stabilizer}. Extending this construction to genuinely multipartite states is nontrivial because the relevant information is encoded in correlations involving all participating qubits. This motivates the development of multipartite ancilla-assisted schemes capable of accessing such global information while preserving the system state.


GHZ-class states constitute one of the most fundamental classes of genuine multipartite entanglement. Unlike Bell states, which characterize bipartite nonlocal correlations between two parties, GHZ states encode global quantum correlations shared simultaneously among all participating qubits. These correlations are intrinsically multipartite and cannot, in general, be reconstructed from independent two-qubit measurements. This distinguishes GHZ-class states from their bipartite Bell-state counterparts and makes their discrimination substantially more demanding. At the same time, the multipartite structure of GHZ states enables important applications, including multipartite quantum secret sharing, distributed quantum computation, conference key agreement, quantum network synchronization, and quantum-enhanced sensing. GHZ states also play a central role in quantum error correction and fault-tolerant quantum computation through their connection with stabilizer and graph-state formalisms. As quantum technologies develop toward distributed quantum networks involving many parties, efficient methods for identifying and certifying multipartite GHZ resources without destroying them become increasingly important. Despite their importance, scalable protocols for the strict nondestructive discrimination of distributed GHZ-class states remain relatively unexplored. Recent progress in distributed quantum computing and networked quantum platforms has further emphasized the importance of generating, certifying, and reusing multipartite entangled resource states across distant nodes \cite{main2025distributed,boschero2025distributed,muralidharan2025simulation}. In this context, the ability to nondestructively identify GHZ-class states shared among multiple parties constitutes an important primitive for scalable quantum architectures.

In this work, we present a scalable ancilla-assisted protocol for the nondestructive discrimination of $n$-qubit GHZ-class states shared among distant parties. By employing multipartite GHZ ancillary resources together with carefully designed local unitary interactions, we show that the global parity and relative phase information of an arbitrary GHZ-class state can be coherently mapped onto two ancillary registers while leaving the system state unchanged. This establishes a scalable distributed framework for strict nondestructive discrimination of multipartite GHZ-class states, showing how global multipartite correlations can be accessed through local operations assisted by preshared entanglement.

Beyond the ideal protocol, we further investigate realistic scenarios in which the ancillary GHZ resources are imperfect. By considering both depolarizing and amplitude-damping noise, we characterize the resulting degradation of the ancillary resources and its effect on the strict nondestructive success probability. For the depolarizing model, we obtain a closed-form success probability and the corresponding critical noise threshold, while the amplitude-damping analysis provides an exact description of the noisy ancillary state and its associated success probability. This analysis provides insight into the practical feasibility of implementing multipartite nondestructive discrimination protocols in near-term quantum network platforms. The noisy-ancilla analysis also suggests an alternative interpretation in terms of decoherence sensing. Since the ideal GHZ ancillary state is known and its degradation can be related analytically to the parameters of the characterized noise models, the ancillary GHZ resources can serve as calibrated multipartite probes of environmental noise \cite{rossi2015entangled,carvalho2004decoherence}. This observation motivates the interpretation of the present framework as a possible quantum decoherence meter, while a detailed optimization of sensing precision and parameter estimation is left for future work.


The remainder of the paper is organized as follows. In Sec.~\ref{sec:protocol}, we introduce the ancilla-assisted nondestructive discrimination protocol for GHZ-class states and explicitly construct the discrimination rules for three- and four-qubit systems, followed by a generalization to arbitrary $n$-qubit GHZ-class states. In Sec.~\ref{sec:noise_analysis}, we investigate the effect of imperfect ancillary resources using both depolarizing and amplitude-damping noise models and we derive a 
expression for the strict nondestructive success probability. 
The same section also discusses the resulting interpretation in terms of decoherence sensing. In Sec.~\ref{sec:experimental_feasibility}, we discuss the experimental implementation of the protocol using the IBM Qiskit simulator. Finally, Sec.~\ref{sec:conclusion} summarizes the main results and discusses possible directions for future work.

\section{Nondestructive quantum state discrimination assisted by ancillary entanglement} 
\label{sec:protocol}


In this section, we present a scalable ancilla-assisted protocol for the strict nondestructive discrimination of multipartite GHZ-class states. The protocol employs two preshared $n$-qubit GHZ ancillary states together with a sequence of local unitary operations that coherently transfer the computational-pattern and relative-phase information of the system GHZ state onto two independent ancillary registers. Projective measurements are performed exclusively on the ancillary qubits, so that the system state is preserved at the end of the complete protocol.


We first illustrate the protocol explicitly for three-qubit GHZ-class states in order to establish the underlying parity--phase mapping mechanism. We then extend the construction to four-qubit GHZ-class states and finally derive a general prescription applicable to arbitrary system size $n$. Since all GHZ-class states considered in this section are pure and the protocol consists entirely of unitary operations, the analysis can be carried out at the level of state vectors. The density-matrix formalism is introduced in Sec.~\ref{sec:noise_analysis}, where noisy ancillary resources are considered.

\subsection{Three-qubit GHZ protocol}
\label{3-qubit protocol1}

\begin{figure}[t]
\centering
\adjustbox{max width=\textwidth}{%
\begin{tikzpicture}[x=1cm,y=1cm,
  every node/.style={scale=0.95},
  rail/.style={line width=0.8pt},
  ctrl/.style={font=\small},
  targ/.style={font=\small},
  gate/.style={draw, rounded corners=3pt, fill=white,
               minimum width=0.6cm, minimum height=0.5cm,
               inner sep=1pt, font=\small},
  state_env/.style={draw, dashed, rounded corners=10pt,
                    fill=gray!5, inner sep=4pt}
]
\node at (3,2.1) {\small Parity};
\node at (5,2.1) {\small Phase};
\node[font=\bfseries, text=blue]   at (6.5,2.3)  {Alice};
\node[font=\bfseries, text=red]     at (6.5,-1.4) {Bob};
\node[font=\bfseries, text=violet] at (6.5,-4.8) {Charlie};
\draw[dotted,thick] (1,1.9) rectangle (11,-0.7);
\draw[dotted,thick] (1,-1.7) rectangle (11,-4.1);
\draw[dotted,thick] (1,-5.1) rectangle (11,-7.5);
\node[text=blue]   at (0.5,1.4)  {$a_1$};
\node[text=blue]   at (0.5,0.6)  {$a_2$};
\node[text=red]     at (0.5,-2.8) {$b_1$};
\node[text=red]     at (0.5,-3.6) {$b_2$};
\node[text=violet] at (0.5,-6.2) {$c_1$};
\node[text=violet] at (0.5,-7.0) {$c_2$};
\foreach \y in {1.4,0.6}   \draw[rail, blue]   (1.0,\y) -- (11.0,\y);
\foreach \y in {-2.8,-3.6} \draw[rail, red]     (1.0,\y) -- (11.0,\y);
\foreach \y in {-6.2,-7.0} \draw[rail, violet] (1.0,\y) -- (11.0,\y);
\draw[rail, blue]   (-0.55,-0.2) -- (12.35,-0.2);
\draw[rail, red]    (-0.55,-2.0) -- (12.35,-2.0);
\draw[rail, violet] (-0.55,-5.4) -- (12.35,-5.4);
\node[state_env, minimum height=6.2cm, minimum width=1.6cm]
  at (-0.9,-2.8) {};
\node[rotate=90, font=\small\bfseries]
  at (-2.15,-2.8) {Input GHZ State};
\node[text=blue]   at (-1.35,-0.2) {$A$};
\node[text=red]    at (-1.35,-2.0) {$B$};
\node[text=violet] at (-1.35,-5.4) {$C$};
\fill[blue]   (-0.55,-0.2) circle (0.12);
\fill[red]    (-0.55,-2.0) circle (0.12);
\fill[violet] (-0.55,-5.4) circle (0.12);
\node[state_env, minimum height=6.2cm, minimum width=1.6cm]
  at (12.7,-2.8) {};
\node[rotate=90, font=\small\bfseries]
  at (13.95,-2.8) {Output GHZ State};
\fill[blue]   (12.35,-0.2) circle (0.12);
\fill[red]    (12.35,-2.0) circle (0.12);
\fill[violet] (12.35,-5.4) circle (0.12);
\node[text=blue]   at (13.15,-0.2) {$A$};
\node[text=red]    at (13.15,-2.0) {$B$};
\node[text=violet] at (13.15,-5.4) {$C$};
\draw (3,-0.2) -- (3,0.6);
\node[ctrl] at (3,-0.2) {$\bullet$};
\node[targ] at (3,0.6) {$\oplus$};
\draw (3,-2.0) -- (3,-3.6);
\node[ctrl] at (3,-2.0) {$\bullet$};
\node[targ] at (3,-3.6) {$\oplus$};
\draw (3,-5.4) -- (3,-7.0);
\node[ctrl] at (3,-5.4) {$\bullet$};
\node[targ] at (3,-7.0) {$\oplus$};
\draw (5,1.4) -- (5,-0.2);
\node[ctrl] at (5,1.4) {$\bullet$};
\node[targ] at (5,-0.2) {$\oplus$};
\draw (5,-2.8) -- (5,-2.0);
\node[ctrl] at (5,-2.8) {$\bullet$};
\node[targ] at (5,-2.0) {$\oplus$};
\draw (5,-6.2) -- (5,-5.4);
\node[ctrl] at (5,-6.2) {$\bullet$};
\node[targ] at (5,-5.4) {$\oplus$};
\node[gate, draw=black, text=black] at (7,1.4)  {$\mathrm{H}$};
\node[gate, draw=black, text=black] at (7,-2.8) {$\mathrm{H}$};
\node[gate, draw=black, text=black] at (7,-6.2) {$\mathrm{H}$};
\node[gate, draw=purple, text=purple] at (10.4,1.4)  {$\mathrm{M}$};
\node[gate, draw=purple, text=purple] at (10.4,0.6)  {$\mathrm{M}$};
\node[gate, draw=purple, text=purple] at (10.4,-2.8) {$\mathrm{M}$};
\node[gate, draw=purple, text=purple] at (10.4,-3.6) {$\mathrm{M}$};
\node[gate, draw=purple, text=purple] at (10.4,-6.2) {$\mathrm{M}$};
\node[gate, draw=purple, text=purple] at (10.4,-7.0) {$\mathrm{M}$};
\end{tikzpicture}%
}
\caption{ \fontsize{9}{12}\selectfont
Quantum circuit for strict nondestructive discrimination of three-qubit
GHZ-class states distributed among Alice ($A$), Bob ($B$), and Charlie ($C$).
Each party's qubit lines are drawn in a separate colour (blue for Alice, red for
Bob, violet for Charlie); the CNOT operations are shown in black, the Hadamard
gates in green, and the measurements in purple. The system qubits $(A,B,C)$
share an unknown GHZ-class state, while two ancillary GHZ states
$(a_1,b_1,c_1)$ and $(a_2,b_2,c_2)$ are preshared among the parties. Local CNOT
operations $A\!\rightarrow a_2$, $B\!\rightarrow b_2$, and $C\!\rightarrow c_2$
transfer the parity information of the system state to the second ancilla.
Subsequently, CNOT operations $a_1\!\rightarrow A$, $b_1\!\rightarrow B$, and
$c_1\!\rightarrow C$ correlate the system with the first ancilla, enabling phase
extraction. Hadamard gates on $(a_1,b_1,c_1)$ convert phase information into
computational-basis correlations. Measurement of the ancillary qubits reveals
the GHZ-class state nondestructively while leaving the system state unchanged.
}
\label{fig:circuit3}
\end{figure}
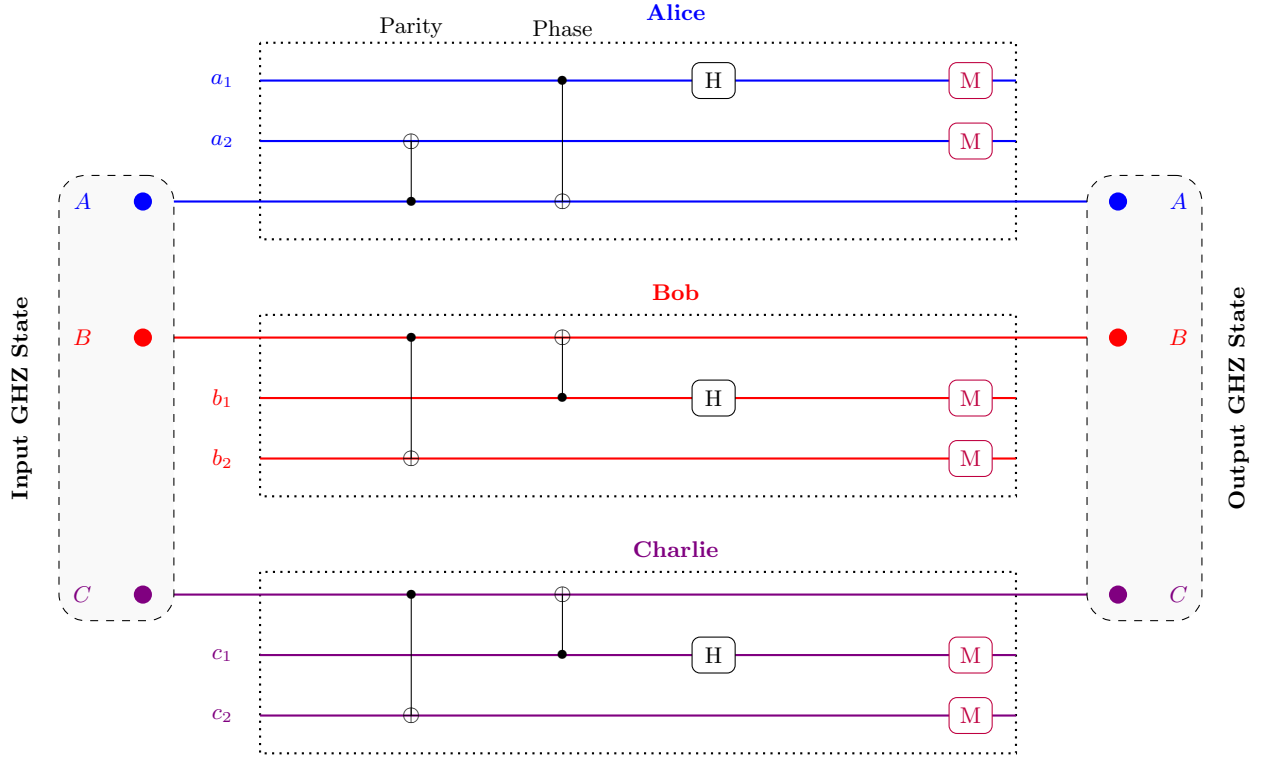

To illustrate the mechanism explicitly, we consider three spatially separated parties:
Alice ($A$), Bob ($B$), and Charlie ($C$), who share an unknown three-qubit GHZ-class state. The three-qubit GHZ-class basis shared by the parties consists of the $2^3=8$ orthogonal states
\begin{equation}
\ket{\psi^{\pm}_{ijk}}_{ABC}
=
\frac{1}{\sqrt{2}}
\left(
\ket{ijk}_{ABC}
\pm
\ket{\bar{i}\bar{j}\bar{k}}_{ABC}
\right), \nonumber
\label{eq:GHZclass}
\end{equation}
where $i,j,k \in \{0,1\}$ and $\bar{i}=1-i$. The subscript $ijk$ specifies the computational-basis pair forming the GHZ superposition. For example,
$\ket{\psi^{+}_{000}}_{ABC} =(\ket{000}_{ABC}+\ket{111}_{ABC})/\sqrt{2}$,
while
$\ket{\psi^{+}_{001}}_{ABC} =(\ket{001}_{ABC}+\ket{110}_{ABC})/\sqrt{2}$. For concreteness, we analyze the representative state
\begin{equation}
\ket{\psi^+_{000}}_{ABC}
=
\frac{1}{\sqrt{2}}
\left(
\ket{000}_{ABC}
+
\ket{111}_{ABC}
\right).
\end{equation}

Each party possesses two ancillary qubits. The first ancillary register $(a_1,b_1,c_1)$ forms one preshared three-qubit GHZ state, while the second ancillary register $(a_2,b_2,c_2)$ forms another identical GHZ state
\begin{equation}
\ket{\phi^+}_{a_1 b_1 c_1}
=
\frac{1}{\sqrt{2}}
\left(
\ket{000}_{a_1 b_1 c_1}
+
\ket{111}_{a_1 b_1 c_1}
\right),
\end{equation}

\begin{equation}
\ket{\phi^+}_{a_2 b_2 c_2}
=
\frac{1}{\sqrt{2}}
\left(
\ket{000}_{a_2 b_2 c_2}
+
\ket{111}_{a_2 b_2 c_2}
\right).
\end{equation}

Uppercase letters $(A,B,C)$ denote system qubits, while lowercase letters $(a_i,b_i,c_i)$ denote ancillary qubits. Throughout this work, $\otimes$ denotes the tensor product between distinct subsystems. For simplicity, tensor products between computational basis states are written in compressed notation, e.g., $\ket{000}_{ABC} = \ket{0}_A \otimes \ket{0}_B \otimes \ket{0}_C$. In what follows, subsystem labels are omitted when no ambiguity arises. In contrast, the ancillary registers are fixed to the reference state $\ket{\phi^+}=(\ket{000}+\ket{111})/\sqrt{2}$, and therefore no additional subscript is required.

The full distributed circuit is shown in Fig.~\ref{fig:circuit3}, where each party performs local operations. No global multi-qubit interaction is required. The total initial state of the system and ancillas is
\begin{equation}
\ket{\Psi^{(0)}}
=
\ket{\phi^+}_{a_1 b_1 c_1}
\otimes
\ket{\phi^+}_{a_2 b_2 c_2}
\otimes
\ket{\psi^+_{000}}_{ABC}.
\end{equation}

For clarity, we first summarize the successive transformations of the joint system–ancilla state. The detailed derivation of each step is presented below.

\[
\ket{\Psi^{(0)}}
\xrightarrow{\substack{\text{Local CNOTs}\\
\text{System}\to\text{Ancilla}_2}}
\ket{\Psi^{(1)}}\\
\xrightarrow{\substack{\text{Local CNOTs}\\
\text{Ancilla}_1\to\text{System}}}
\ket{\Psi^{(2)}}
\xrightarrow{\substack{H^{\otimes3}\\
\text{on }\text{Ancilla}_1}}
\ket{\Psi^{(3)}}.
\]

The first transformation, $\ket{\Psi^{(0)}}\rightarrow\ket{\Psi^{(1)}}$, is implemented by the local controlled-NOT (CNOT) operations applied between the system qubits and the second ancillary register. Specifically, the operations
\[
\mathrm{CNOT}_{A\rightarrow a_2}, \quad
\mathrm{CNOT}_{B\rightarrow b_2}, \quad
\mathrm{CNOT}_{C\rightarrow c_2},
\]
transfer the parity information of the system onto the second ancilla. Here $\mathrm{CNOT}_{i\rightarrow j}$ denotes a local CNOT gate with qubit $i$ as control and qubit $j$ as target.

For the symmetric input $\ket{\psi^+_{000}}_{ABC}$, both computational components $\ket{000}$ and $\ket{111}$ induce identical collective flips on $\ket{\phi^+}_{a_2 b_2 c_2}$. Since
\[
\sigma_X^{\otimes 3}\ket{\phi^+}=\ket{\phi^+},
\]
where $\sigma_X$ denotes the Pauli-$X$ operator, the second ancilla is invariant under collective flips and the total state remains unchanged,
\begin{equation}
\ket{\Psi^{(1)}}=\ket{\Psi^{(0)}}.
\end{equation}

For a general GHZ-class input, however, this operation imprints the parity pattern onto the second ancillary register.

The second transformation, $\ket{\Psi^{(1)}}\rightarrow\ket{\Psi^{(2)}}$,
is implemented by subsequently applying local CNOT gates
\[
\mathrm{CNOT}_{a_1\rightarrow A}, \quad
\mathrm{CNOT}_{b_1\rightarrow B}, \quad
\mathrm{CNOT}_{c_1\rightarrow C},
\]
which correlate the system qubits with the first ancillary register. For the symmetric GHZ input state, both computational components $\ket{000}$ and $\ket{111}$ induce identical collective bit flips on the ancilla–system pair. Since the GHZ states are invariant under collective Pauli-$X$ operations,
\[
\sigma_X^{\otimes3}\ket{\phi^+} = \ket{\phi^+}, 
\qquad
\sigma_X^{\otimes3}\ket{\psi^+_{000}} = \ket{\psi^+_{000}},
\]
 the joint state of the first ancilla and the system remains unchanged under these collective CNOT operations. Consequently,
\begin{equation}
\ket{\Psi^{(2)}}=\ket{\Psi^{(1)}}.
\end{equation}

Finally, the last transformation,
$\ket{\Psi^{(2)}}\rightarrow\ket{\Psi^{(3)}}$,
results from applying Hadamard gates to the first ancillary register
$(a_1,b_1,c_1)$. These gates convert the global phase information into computational-basis correlations. Here $H$ denotes the single-qubit Hadamard operator. Using
\begin{equation}
H^{\otimes3}\ket{\phi^+}
=
\frac{1}{2}
\left(
\ket{000}+\ket{011}+\ket{101}+\ket{110}
\right), \nonumber
\end{equation}
we see that the symmetric GHZ ancilla produces  even-parity computational strings.

More generally, for an arbitrary GHZ-class state we use
\begin{align}
H^{\otimes 3}\ket{000}
&=
\frac{1}{\sqrt{8}}
\sum_{x\in\{0,1\}^3} \ket{x}, \nonumber\\
H^{\otimes 3}\ket{111}
&=
\frac{1}{\sqrt{8}}
\sum_{x\in\{0,1\}^3} (-1)^{w(x)} \ket{x}, \nonumber
\end{align}
where $x=(x_1,x_2,x_3)$ denotes a three-bit computational basis string and $w(x)=x_1+x_2+x_3$ is its Hamming weight, i.e., the number of ones in the string.

Consequently,
\begin{equation}
H^{\otimes3}
\frac{1}{\sqrt{2}}
\left(
\ket{000}\pm\ket{111}
\right)
=
\frac{1}{4}
\sum_{x\in\{0,1\}^3}
\left(
1\pm(-1)^{w(x)}
\right)
\ket{x}. \nonumber
\end{equation}

Therefore, for the $+$ state, only even-parity strings (those with $w(x)$ even) survive, while for the $-$ state, only odd-parity strings survive. Thus, for the representative state $\ket{\psi^+_{000}}_{ABC}$, the two layers of CNOT operations preserve the GHZ structure due to the symmetry of the ancillary states under collective bit flips, while the Hadamard layer transfers the global phase information to the parity sector of the first ancilla.

For the input $\ket{\psi^+_{000}}_{ABC}$, the joint state becomes

\begin{equation}
  \ket{\Psi^{(3)}} =
\frac{1}{2}
\left(
\ket{000}+\ket{011}+\ket{101}+\ket{110}
\right)_{a_1 b_1 c_1}
\otimes
\ket{\phi^+}_{a_2 b_2 c_2}
\otimes
\ket{\psi^+_{000}}_{ABC}.  
\end{equation}

Although the above derivation was presented for the representative state
$\ket{\psi^+_{000}}_{ABC}$, the protocol applies identically to any GHZ-class basis state. For a general input $\ket{\psi^{\pm}_{ijk}}_{ABC}$,
the final joint state is
\begin{equation}
\ket{\Psi^{(3)}} =
\frac{1}{2}
\sum_{x\in\mathcal{P}_{\pm}}
\ket{x}_{a_1 b_1 c_1}
\otimes
\ket{\psi^{\pm}_{ijk}}_{a_2 b_2 c_2}
\otimes
\ket{\psi^{\pm}_{ijk}}_{ABC},
\end{equation}
where $\mathcal{P}_+$ ($\mathcal{P}_-$) denotes the set of even-parity (odd-parity) three-bit strings of the first ancillary register. Thus, the second ancillary register coherently records the computational bit pattern $(ijk)$ of the GHZ superposition, while the first ancillary register encodes the global phase through its parity sector.

\begin{table}[t]
\centering
\caption{ \fontsize{9}{12}\selectfont
Allowed ancillary measurement outcomes for strict nondestructive discrimination of three-qubit GHZ-class states $\ket{\psi^{\pm}_{ijk}}_{ABC}$. Both ancillary registers are initialized in $\ket{\phi^+}=(\ket{000}+\ket{111})/\sqrt2$. For each input state, exactly eight computational-basis ancillary outcomes occur with nonzero probability, while the system state remains unchanged throughout the protocol.}
\label{tab:3qubit_GHZ}
\renewcommand{\arraystretch}{1.25}

\begin{tabular}{@{}p{0.30\textwidth}p{0.65\textwidth}@{}}
\hline\hline
\textbf{\rule[-2.2ex]{0pt}{3.6ex}System state $\ket{\psi^{\pm}_{ijk}}_{ABC}$}
& \textbf{Allowed ancillary outcomes $(a_1 b_1 c_1 a_2 b_2 c_2)$} \\
\hline
$\ket{\psi^+_{000}}=\frac{1}{\sqrt2}(\ket{000}+\ket{111})$
& $000000,\;000111,\;011000,\;011111,\;101000,\;101111,\;110000,\;110111$ \\

$\ket{\psi^-_{000}}=\frac{1}{\sqrt2}(\ket{000}-\ket{111})$
& $001000,\;001111,\;010000,\;010111,\;100000,\;100111,\;111000,\;111111$ \\

$\ket{\psi^+_{001}}=\frac{1}{\sqrt2}(\ket{001}+\ket{110})$
& $000001,\;000110,\;011001,\;011110,\;101001,\;101110,\;110001,\;110110$ \\

$\ket{\psi^-_{001}}=\frac{1}{\sqrt2}(\ket{001}-\ket{110})$
& $001001,\;001110,\;010001,\;010110,\;100001,\;100110,\;111001,\;111110$ \\

$\ket{\psi^+_{010}}=\frac{1}{\sqrt2}(\ket{010}+\ket{101})$
& $000010,\;000101,\;011010,\;011101,\;101010,\;101101,\;110010,\;110101$ \\

$\ket{\psi^-_{010}}=\frac{1}{\sqrt2}(\ket{010}-\ket{101})$
& $001010,\;001101,\;010010,\;010101,\;100010,\;100101,\;111010,\;111101$ \\

$\ket{\psi^+_{011}}=\frac{1}{\sqrt2}(\ket{011}+\ket{100})$
& $000011,\;000100,\;011011,\;011100,\;101011,\;101100,\;110011,\;110100$ \\

$\ket{\psi^-_{011}}=\frac{1}{\sqrt2}(\ket{011}-\ket{100})$
& $001011,\;001100,\;010011,\;010100,\;100011,\;100100,\;111011,\;111100$ \\

\hline\hline
\end{tabular}
\end{table}

Projective measurements in the computational basis are then performed exclusively on the ancillary qubits $(a_1 b_1 c_1 a_2 b_2 c_2)$, while the system qubits $(A,B,C)$ remain unmeasured. Throughout this work we use the term strict nondestructive discrimination to denote the situation in which the measurement outcomes identify the GHZ-class state while leaving the system qubits exactly in the original state, i.e., the reduced system state is unchanged after the protocol. For the input $\ket{\psi^+_{000}}_{ABC}$, exactly eight
bit strings
\[\{
000000,\,000111,\,011000,\,011111,
101000,\,101111,\,110000,\,110111
\},
\]
occur with nonzero probability, while all remaining $2^6-8$ outcomes vanish identically. The complete set of allowed ancillary outcomes for all GHZ-class inputs is listed in Table~\ref{tab:3qubit_GHZ}. These outcomes correspond to the combinations of the four even-parity strings produced by the first ancillary register and the two computational patterns encoded in the second ancilla.

The protocol separates classical parity information and relative phase information into two independent ancillary registers. The second ancillary GHZ state coherently records the computational-basis structure of the GHZ superposition through local CNOT operations, while the first ancillary register, after the Hadamard transformation, converts the global phase into measurable parity correlations. Consequently, the complete GHZ label is transferred coherently to the ancillary degrees of freedom.

Since the final state factorizes between system and ancillas and no measurement is performed on the system qubits $(A,B,C)$, the reduced state of the system qubits remains identical to the input GHZ state. The discrimination is therefore strictly nondestructive: the system is neither projected nor disturbed and can be reused for subsequent quantum information processing tasks. 

Compared to the bipartite Bell-state case, where only two correlated ancillary outcomes arise, the three-qubit GHZ protocol yields $2^3$ valid ancillary patterns for each GHZ-class state. This increase reflects the richer combinatorial structure of multipartite parity correlations. Nevertheless, the logical structure of the protocol remains unchanged: parity information is extracted first, followed by phase information.

The structure revealed in the three-qubit case directly generalizes to larger multipartite systems. In particular, the same sequence of local CNOT operations and Hadamard transformations separates the computational pattern and the global phase information into two ancillary GHZ registers. We now illustrate this extension explicitly for four-qubit GHZ-class states, which already exhibits the general features of the $n$-qubit protocol.

\subsection{Four-qubit GHZ protocol}

\begin{table}[t]
\centering
\caption{ \fontsize{9}{12}\selectfont
Allowed ancillary measurement outcomes for strict nondestructive discrimination of four-qubit GHZ-class states. For $+$ states the first ancillary register yields even-parity strings $\{0000,0011,0101,0110,1001,1010,1100,1111\}$, while for $-$ states it yields odd-parity strings $\{0001,0010,0100,0111,1000,1011,1101,1110\}$. The symbol $\oplus$ denotes the concatenation of the two bit-string sets associated with the first and second ancillary registers, producing $8 \times 2 = 16$ distinct eight-bit outcomes for each input state. For example, for the input state 
$\ket{\psi^+_{0000}}=\frac{1}{\sqrt2}(\ket{0000}+\ket{1111})$, 
the allowed eight-bit ancillary outcomes are $\{00000000, 00001111, 00110000, 00111111, 01010000, 01011111, 01100000, 01101111, 10010000, 10011111, \newline 10100000, 10101111, 11000000, 11001111, 11110000,11111111\}$, corresponding to the combinations of the first-register even-parity strings $\{0000,0011,0101,0110,1001,1010,1100,1111\}$ with the second-register strings
$\{0000,1111\}$. Both ancillary registers are initialized in $\ket{\phi^+}=(\ket{0000}+\ket{1111})/\sqrt2$. Out of $2^8=256$ possible outcomes, exactly $2^4=16$ occur for each GHZ-class state while the system state remains unchanged.
}
\label{tab:4qubit__GHZ}
\renewcommand{\arraystretch}{1.25}

\begin{tabular}{@{}p{0.38\linewidth}p{0.58\linewidth}@{}}
\hline\hline
\textbf{\rule[-2.2ex]{0pt}{3.6ex}System state $\ket{\psi^{\pm}_{ijkl}}_{ABCD}$}
& \textbf{Ancillary outcomes $(a_1 b_1 c_1 d_1\, a_2 b_2 c_2 d_2)$} \\
\hline

$\ket{\psi^+_{0000}}=\frac{1}{\sqrt2}(\ket{0000}+\ket{1111})$
& $\{0000,0011,0101,0110,1001,1010,1100,1111\} \oplus \{0000,1111\}$ \\

$\ket{\psi^-_{0000}}=\frac{1}{\sqrt2}(\ket{0000}-\ket{1111})$
& $\{0001,0010,0100,0111,1000,1011,1101,1110\} \oplus \{0000,1111\}$ \\[4pt]

$\ket{\psi^+_{0001}}=\frac{1}{\sqrt2}(\ket{0001}+\ket{1110})$
& $\{0000,0011,0101,0110,1001,1010,1100,1111\} \oplus \{0001,1110\}$ \\

$\ket{\psi^-_{0001}}=\frac{1}{\sqrt2}(\ket{0001}-\ket{1110})$
& $\{0001,0010,0100,0111,1000,1011,1101,1110\} \oplus \{0001,1110\}$ \\[4pt]

$\ket{\psi^+_{0010}}=\frac{1}{\sqrt2}(\ket{0010}+\ket{1101})$
& $\{0000,0011,0101,0110,1001,1010,1100,1111\} \oplus \{0010,1101\}$ \\

$\ket{\psi^-_{0010}}=\frac{1}{\sqrt2}(\ket{0010}-\ket{1101})$
& $\{0001,0010,0100,0111,1000,1011,1101,1110\} \oplus \{0010,1101\}$ \\[4pt]

$\ket{\psi^+_{0011}}=\frac{1}{\sqrt2}(\ket{0011}+\ket{1100})$
& $\{0000,0011,0101,0110,1001,1010,1100,1111\} \oplus \{0011,1100\}$ \\

$\ket{\psi^-_{0011}}=\frac{1}{\sqrt2}(\ket{0011}-\ket{1100})$
& $\{0001,0010,0100,0111,1000,1011,1101,1110\} \oplus \{0011,1100\}$ \\[4pt]

$\ket{\psi^+_{0100}}=\frac{1}{\sqrt2}(\ket{0100}+\ket{1011})$
& $\{0000,0011,0101,0110,1001,1010,1100,1111\} \oplus \{0100,1011\}$ \\

$\ket{\psi^-_{0100}}=\frac{1}{\sqrt2}(\ket{0100}-\ket{1011})$
& $\{0001,0010,0100,0111,1000,1011,1101,1110\} \oplus \{0100,1011\}$ \\[4pt]

$\ket{\psi^+_{0101}}=\frac{1}{\sqrt2}(\ket{0101}+\ket{1010})$
& $\{0000,0011,0101,0110,1001,1010,1100,1111\} \oplus \{0101,1010\}$ \\

$\ket{\psi^-_{0101}}=\frac{1}{\sqrt2}(\ket{0101}-\ket{1010})$
& $\{0001,0010,0100,0111,1000,1011,1101,1110\} \oplus \{0101,1010\}$ \\[4pt]

$\ket{\psi^+_{0110}}=\frac{1}{\sqrt2}(\ket{0110}+\ket{1001})$
& $\{0000,0011,0101,0110,1001,1010,1100,1111\} \oplus \{0110,1001\}$ \\

$\ket{\psi^-_{0110}}=\frac{1}{\sqrt2}(\ket{0110}-\ket{1001})$
& $\{0001,0010,0100,0111,1000,1011,1101,1110\} \oplus \{0110,1001\}$ \\[4pt]

$\ket{\psi^+_{0111}}=\frac{1}{\sqrt2}(\ket{0111}+\ket{1000})$
& $\{0000,0011,0101,0110,1001,1010,1100,1111\} \oplus \{0111,1000\}$ \\

$\ket{\psi^-_{0111}}=\frac{1}{\sqrt2}(\ket{0111}-\ket{1000})$
& $\{0001,0010,0100,0111,1000,1011,1101,1110\} \oplus \{0111,1000\}$ \\

\hline\hline
\end{tabular}
\end{table}

We now extend the nondestructive discrimination scheme to four spatially separated parties: Alice ($A$), Bob ($B$), Charlie ($C$), and David ($D$), who share an unknown four-qubit GHZ-class state $\ket{\Psi}_{ABCD}$. The four-qubit GHZ-class basis consists of the $2^4=16$ orthogonal states
\begin{equation}
\ket{\psi^{\pm}_{ijkl}}_{ABCD}
=
\frac{1}{\sqrt{2}}
\left(
\ket{ijkl}_{ABCD}
\pm
\ket{\bar i \bar j \bar k \bar l}_{ABCD}
\right), \nonumber
\end{equation}
where $i,j,k,l\in\{0,1\}$ and $\bar i = 1-i$. For concreteness, we first analyze the representative state
\begin{equation}
\ket{\psi^+_{0000}}_{ABCD}
=
\frac{1}{\sqrt{2}}
\left(
\ket{0000}_{ABCD}
+
\ket{1111}_{ABCD}
\right). 
\end{equation}

Each party possesses two ancillary qubits. The first ancillary register $(a_1,b_1,c_1,d_1)$ forms one preshared four-qubit GHZ state, while the second ancillary register $(a_2,b_2,c_2,d_2)$ forms another identical GHZ state.
\begin{equation}
\ket{\phi^+}_{a_1 b_1 c_1 d_1}
=
\frac{1}{\sqrt{2}}
\left(
\ket{0000}
+
\ket{1111}
\right),
\end{equation}
\begin{equation}
\ket{\phi^+}_{a_2 b_2 c_2 d_2}
=
\frac{1}{\sqrt{2}}
\left(
\ket{0000}
+
\ket{1111}
\right).
\end{equation}

The total initial state of the system and ancillas is
\begin{equation}
\ket{\Psi^{(0)}}
=
\ket{\phi^+}_{a_1 b_1 c_1 d_1}
\otimes
\ket{\phi^+}_{a_2 b_2 c_2 d_2}
\otimes
\ket{\psi^+_{0000}}_{ABCD}.
\end{equation}

Similarly to the 3-qubit case, we first summarise the successive transformations of the joint system–ancilla state. The detailed derivation of each step is presented below.
\[
\ket{\Psi^{(0)}}
\xrightarrow{\substack{\text{Local CNOTs}\\
\text{System}\to\text{Ancilla}_2}}
\ket{\Psi^{(1)}}
\xrightarrow{\substack{\text{Local CNOTs}\\
\text{Ancilla}_1\to\text{System}}}
\ket{\Psi^{(2)}}
\xrightarrow{\substack{H^{\otimes4}\\
\text{on }\text{Ancilla}_1}}
\ket{\Psi^{(3)}}.
\]
The first transformation, $\ket{\Psi^{(0)}}\rightarrow\ket{\Psi^{(1)}}$, is generated by a layer of local CNOT gates applied between the system qubits and the second ancillary register. Specifically,
\[
\mathrm{CNOT}_{A\rightarrow a_2},\quad
\mathrm{CNOT}_{B\rightarrow b_2},\quad
\mathrm{CNOT}_{C\rightarrow c_2},\quad
\mathrm{CNOT}_{D\rightarrow d_2},
\]
which imprint the computational pattern (parity structure) of the system onto the second ancillary register. For the symmetric input $\ket{\psi^+_{0000}}_{ABCD}$, the components $\ket{0000}$ and $\ket{1111}$ induce identical collective flips on $\ket{\phi^+}_{a_2 b_2 c_2 d_2}$. 
Since
\[
\sigma_X^{\otimes4}\ket{\phi^+}=\ket{\phi^+},
\]
the second ancillary GHZ state is invariant under collective flips, and therefore the global state remains unchanged,
\begin{equation}
\ket{\Psi^{(1)}}=\ket{\Psi^{(0)}}.
\end{equation}

For a general four-qubit GHZ-class input, however, this operation imprints the computational parity structure of the system onto the second ancillary register.

The second transformation, $\ket{\Psi^{(1)}}\rightarrow\ket{\Psi^{(2)}}$,
is generated by subsequently applying a second layer of local CNOT gates
\[
\mathrm{CNOT}_{a_1\rightarrow A},\quad
\mathrm{CNOT}_{b_1\rightarrow B},\quad
\mathrm{CNOT}_{c_1\rightarrow C},\quad
\mathrm{CNOT}_{d_1\rightarrow D},
\]
correlates the system qubits with the first ancillary register. Since the four-qubit GHZ state is invariant under collective bit flips, namely
\[
\sigma_X^{\otimes4}\ket{\phi^+}=\ket{\phi^+},
\qquad
\sigma_X^{\otimes4}\ket{\psi^+_{0000}}=\ket{\psi^+_{0000}},
\]
the joint state of the first ancilla and the system remains unchanged under these collective CNOT operations. Consequently,
\begin{equation}
\ket{\Psi^{(2)}}=\ket{\Psi^{(1)}}.
\end{equation}

Finally, the last transformation,
$\ket{\Psi^{(2)}}\rightarrow\ket{\Psi^{(3)}}$,
is obtained by applying Hadamard gates to each qubit of the first ancillary register $(a_1,b_1,c_1,d_1)$. 
Using similar procedure in Sec.~\ref{3-qubit protocol1} we can obtain
\begin{equation}
H^{\otimes4}
\frac{1}{\sqrt2}
(\ket{0000}\pm\ket{1111})
=
\frac{1}{4\sqrt2}
\sum_{x\in\{0,1\}^4}
\left(
1 \pm (-1)^{w(x)}
\right)
\ket{x}. \nonumber
\end{equation}

For the representative state $\ket{\psi^+_{0000}}$, only computational-basis strings with even $w(x)$  survive. The joint state therefore becomes 
\begin{equation}
\ket{\Psi^{(3)}}
=
\frac{1}{\sqrt{8}}
\sum_{x\in\mathcal{P}_+}
\ket{x}_{a_1 b_1 c_1 d_1}
\otimes
\ket{\phi^+}_{a_2 b_2 c_2 d_2}
\otimes
\ket{\psi^+_{0000}}_{ABCD},
\end{equation}.

Although the above derivation illustrates the protocol for the representative state $\ket{\psi^+_{0000}}_{ABCD}$, the protocol applies identically to any GHZ-class basis state.
For a general input
$\ket{\psi^{\pm}_{ijkl}}_{ABCD}$, the same sequence of local operations yields
\begin{equation}
\ket{\Psi^{(3)}}
=
\frac{1}{\sqrt{8}}
\sum_{x\in\mathcal{P}_{\pm}}
\ket{x}_{a_1 b_1 c_1 d_1}
\otimes
\ket{\psi^{\pm}_{ijkl}}_{a_2 b_2 c_2 d_2}
\otimes
\ket{\psi^{\pm}_{ijkl}}_{ABCD}.
\end{equation}

Thus, the second ancillary register remains in the GHZ-class state corresponding to the computational pattern $(ijkl)$, coherently encoding the classical pattern $(ijkl)$ without disturbing the entanglement of the system. Meanwhile, the first ancillary register encodes the global phase information through its parity sector. The final state factorizes between the ancillary registers and the system qubits, ensuring that the GHZ-class state of $(A,B,C,D)$ is preserved throughout the protocol.

Projective measurements are performed exclusively on the ancillary qubits $(a_1 b_1 c_1 d_1 a_2 b_2 c_2 d_2)$. Out of the $2^8=256$ possible outcomes, exactly $2^4=16$ correlated eight-bit strings occur for each GHZ-class state, as listed in Table \ref{tab:4qubit__GHZ}. Because the final state factorizes between the ancillary registers and the system, and the system qubits are never measured, the state $\ket{\psi^{\pm}_{ijkl}}_{ABCD}$ remains unchanged.

Compared to the three-qubit case, where $2^3=8$ valid ancillary patterns appear, the four-qubit protocol yields $2^4=16$ correlated outcomes per GHZ-class state. The logical structure is identical to the three-qubit protocol: the second ancillary register encodes the computational pattern,
while the first ancillary register converts relative phase into parity correlations. The increase from $2^3$ to $2^4$ valid patterns reflects only the enlarged multipartite parity structure.


\subsection{Generalization to arbitrary \texorpdfstring{$n$}{n}-qubit GHZ-class states}

The nondestructive discrimination protocol developed for the three- and four-qubit cases extends naturally to an arbitrary $n$-qubit GHZ-class state. The GHZ-class basis for $n$ qubits consists of the $2^n$ orthogonal states
\begin{equation}
\ket{\psi^{\pm}_{b}}
=
\frac{1}{\sqrt{2}}
\left(
\ket{b}
\pm
\ket{\bar b}
\right), \nonumber
\end{equation}
where $b=b_1 b_2 \dots b_n \in \{0,1\}^n$ is an $n$-bit computational string and $\bar b$ denotes its bitwise complement.

The generalized protocol employs two ancillary $n$-qubit GHZ states together with local CNOT operations acting pairwise between corresponding system and ancillary qubits. In the first layer, CNOT gates
$\mathrm{CNOT}_{S_i\rightarrow \text{anc}_{2,i}}$ (with the system qubits $S_i$ as controls and the qubits of the second ancillary register $\text{anc}_{2,i}$ as targets) coherently encode the computational pattern $b$ onto the second ancilla. In the second layer, CNOT gates $\mathrm{CNOT}_{\text{anc}_{1,i}\rightarrow S_i}$ (with the qubits of the first ancillary register $\text{anc}_{1,i}$ as controls and  $S_i$ as targets) 
correlate the system qubits with the first ancillary register and transfer the phase information to that ancilla.

These operations do not disturb the system state due to the invariance of GHZ states under collective bit flips,
\begin{equation}
\sigma_X^{\otimes n}\ket{\phi^+}=\ket{\phi^+},
\qquad
\sigma_X^{\otimes n}\ket{\psi^{\pm}_{b}}=\ket{\psi^{\pm}_{b}}. \nonumber
\end{equation}

A subsequent Hadamard transformation applied to all qubits of the first ancillary register converts the relative phase ($+$ or $-$) of the GHZ superposition into parity correlations of the computational-basis strings produced in that register.

Consequently, projective measurements performed exclusively on the $2n$ ancillary qubits yield correlated $2n$-bit computational-basis strings that uniquely determine both the bit string $b$ and the relative phase sign. The system qubits remain in the original state $\ket{\psi^{\pm}_{b}}$, thereby ensuring strict nondestructiveness.

For each input GHZ-class state, only $2^n$ of the total $2^{2n}$ possible ancillary outcomes occur with nonzero probability. These allowed outcomes obey a structured parity-selection rule. Let $x=(x_1,x_2,\dots,x_n)$ denote the $n$-bit measurement outcome of the first ancillary register, and let $w(x)=\sum_{i=1}^n x_i$ be its Hamming weight. The computational component $b$ (or $\bar b$) is encoded in the second ancilla, while the first ancilla produces bit strings whose Hamming weight $w(x)$ is even for the $+$ phase and odd for the $-$ phase.

Equivalently, the allowed ancillary outcomes are strings of the form $x\,b$ and $x\,\bar b$, where the first $n$ bits correspond to the measurement outcome $x$ of the first ancillary register and the remaining $n$ bits correspond to the outcome of the second ancilla.

This construction therefore establishes a scalable, symmetry-based, and fully nondestructive discrimination protocol for arbitrary multipartite GHZ-class states. The central design principle underlying the proposed protocol is the separation of multipartite information into complementary computational-pattern and global-phase sectors, each extracted coherently by an independent ancillary GHZ register. This parity--phase decomposition remains unchanged for arbitrary system size and therefore provides the fundamental mechanism underlying the scalability of the protocol.

The analysis of the $3$-, $4$-, and general $n$-qubit cases shows that the protocol exploits a common structural feature of GHZ-class states: the superposition of two complementary computational strings together with a well-defined global parity structure. Because the protocol relies on these symmetry properties to transfer the computational pattern and phase information to ancillary registers without disturbing the system, it is important to clarify the class of states for which such a strategy remains valid.


\subsection{Scope of applicability}

The above protocol relies crucially on the symmetry structure of GHZ-class states. In particular, GHZ states are superpositions of two computational basis states that are related by a global bit flip, which gives rise to well-defined parity correlations across all qubits \cite{greenberger1989going,greenberger1990bell}. Because of this structure, the protocol can coherently transfer both the computational pattern and the global phase information of the system state onto two ancillary registers while leaving the system itself unchanged.

This mechanism relies on a distinctive property of GHZ-class states: the entire multipartite entanglement is encoded in a coherent superposition of two complementary computational basis strings together with a single global relative phase. This global parity--phase structure is absent in other inequivalent multipartite entanglement classes, such as W states \cite{dur2000three}, whose entanglement is distributed over many computational basis components. Consequently, the present ancilla-assisted strategy is intrinsically tailored to GHZ-class entanglement.

An important feature of the proposed protocol is that the discrimination information is naturally decomposed into two complementary channels. The computational-basis information of the GHZ-class state is coherently transferred to the second ancillary register, whereas the global phase information is encoded in the parity sector of the first ancillary register through the action of Hadamard gates. This separation reflects the underlying stabilizer structure of GHZ-class states and remains valid for arbitrary system size~$n$. Consequently, the protocol admits a scalable distributed implementation in which the same sequence of local operations is applied irrespective of the number of parties, with only the size of the ancillary registers increasing linearly with~$n$.

Although the protocol exploits multipartite GHZ entanglement, an important question remains whether the same task could instead be accomplished using local operations and classical communication (LOCC). We address this question in the following subsection.

\subsection{LOCC limitation for GHZ-class nondestructive discrimination}

Complete nondestructive discrimination of multipartite GHZ-class states shared among spatially separated parties cannot be achieved using only LOCC assisted by separable (non-entangled) ancillary systems \cite{bennett1999quantum,ghosh2001distinguishability}. In particular, any protocol restricted to LOCC with separable ancillas cannot extract both the global parity correlations and the relative phase of a GHZ-class state while simultaneously preserving the state itself.

The origin of this limitation lies in the intrinsically global structure of GHZ-class entanglement. GHZ states are stabilized by genuinely multipartite operators involving all parties collectively, such as $X^{\otimes n}$ \cite{gottesman1997stabilizer}. The logical information distinguishing different GHZ-class states is therefore encoded in nonlocal correlations that cannot be decomposed into purely bipartite contributions.

Consequently, any LOCC protocol employing only separable ancillary systems can access at most locally available or bipartite correlations and thus cannot faithfully extract the global phase–parity structure of the GHZ state without inducing disturbance \cite{horodecki2009quantum}. Measurement of the relevant multipartite stabilizers therefore requires either preshared entangled ancillary resources, as employed in the present protocol, or coherent global joint operations across all parties.

This observation extends the LOCC limitations previously identified for nondestructive Bell-state discrimination, where the achievable success probability under LOCC without entangled ancillas is strictly limited~\cite{ghosh2001distinguishability,NondestructivePRA2025}. In the multipartite GHZ setting, the restriction is even stronger because the distinguishing information resides in genuinely global observables that are inaccessible through LOCC with separable ancillas alone.

Unlike the Bell-state case, where the relevant stabilizer information is encoded in bipartite parity and phase observables, the multipartite GHZ protocol requires simultaneous access to genuinely multipartite correlations shared collectively among all parties. The two ancillary GHZ registers employed in the present protocol provide complementary multipartite resources: the second ancillary register coherently records the computational pattern of the GHZ-class state, whereas the first ancillary register converts the global phase information into measurable parity correlations through local Hadamard operations. Consequently, the protocol demonstrates that the inability of LOCC originates not merely from the absence of entanglement, but from the inability of separable ancillary systems to coherently mediate the extraction of genuinely multipartite parity--phase information while preserving the distributed GHZ state. This provides an operational interpretation of multipartite GHZ entanglement as a resource for nondestructive information extraction rather than merely as the state being discriminated.

From this perspective, the ancillary GHZ states should be viewed as operational resources that enable distributed access to multipartite stabilizer information without requiring global joint measurements. The protocol therefore highlights the essential nonlocal resource required for strict nondestructive discrimination within the present framework: local unitary operations together with preshared multipartite GHZ entanglement. The above discussion establishes the fundamental operational limitations associated with nondestructive GHZ-class discrimination under LOCC constraints. Having clarified these conceptual restrictions, it is important to examine how the protocol performs under realistic physical conditions where ancillary entangled resources are imperfect. We therefore now analyze the robustness of the proposed discrimination scheme in the presence of noise affecting the ancillary GHZ states.


\section{Nondestructive discrimination with noisy GHZ ancillary states}
\label{sec:noise_analysis}


In realistic quantum networks, preshared multipartite entangled resources are inevitably affected by decoherence and operational imperfections. Consequently, the ancillary GHZ states employed by the proposed discrimination protocol cannot, in general, be prepared as ideal maximally entangled states. To quantify the resulting performance degradation, we investigate the robustness of the protocol in the presence of noisy ancillary GHZ resources. Specifically, we consider both depolarizing and amplitude-damping noise acting on the ancillary GHZ states. For the depolarizing channel, we derive 
analytical expressions for the ancillary fidelity, the strict nondestructive success probability, and the corresponding critical noise threshold that marks the quantum advantage over the optimal classical strategy. For the amplitude-damping channel, we derive the exact noisy ancillary GHZ state together with its fidelity, providing an analytical characterization of how this non-unital noise channel degrades the multipartite ancillary resource. Together, these results reveal how the robustness of the proposed protocol depends on both the underlying noise mechanism and the number of qubits.
 
\subsection{Depolarized \texorpdfstring{$n$}{n}-qubit GHZ ancillary state}


We first consider a depolarizing noise model for the ancillary GHZ resources. This model is particularly convenient because it preserves the symmetry of the GHZ basis and allows the quality of the noisy resource to be characterized analytically. It therefore provides a useful benchmark for assessing the robustness of the proposed discrimination protocol.
Throughout this section we take the reference state to be
$|{\rm GHZ}^+\rangle\equiv\ket{\psi^+_{0^n}}$, which corresponds to the GHZ-class state with bit string $b=0^n$. For notational simplicity, we use the standard $|{\rm GHZ}^+\rangle$ notation when referring to the ancillary resource throughout this section.

The noisy ancillary resource is therefore described by the depolarized GHZ-class state
\begin{equation}
W_n(\lambda)
= (1-\lambda)\,|{\rm GHZ}^+\rangle\langle{\rm GHZ}^+|
+ \lambda\,\frac{\mathbb{I}_{2^n}}{2^n},
\label{eq:WernerGHZ}
\end{equation}
where
\begin{equation}
|{\rm GHZ}^+\rangle
\equiv
\ket{\psi^{+}_{0^n}}
=
\frac{1}{\sqrt{2}}
\left(|0\rangle^{\otimes n}+|1\rangle^{\otimes n}\right), \nonumber
\end{equation}
$\mathbb{I}_{2^n}$ denotes the identity operator on the $2^n$-dimensional Hilbert space, and $\lambda\in[0,1]$ quantifies the depolarizing noise strength.
The limits $\lambda=0$ and $\lambda=1$ correspond to a pure GHZ-class state and a maximally mixed state, respectively.

To determine the fidelity of the noisy ancillary resource, we first express the maximally mixed state in an orthonormal basis containing the reference GHZ state $|{\rm GHZ}^{+}\rangle$. Since every normalized state can be extended to a complete orthonormal basis of the Hilbert space, let
\[
\left\{
|{\rm GHZ}^{+}\rangle,
|\phi_2\rangle,
|\phi_3\rangle,
\ldots,
|\phi_{2^n}\rangle
\right\}
\]
be a complete orthonormal basis of the $2^n$-dimensional Hilbert space, where
\[
\langle {\rm GHZ}^{+}|\phi_j\rangle = 0,
\qquad
j=2,\ldots,2^n.
\]
For concreteness, one may choose $\{|\phi_j\rangle\}$ to be the remaining orthonormal GHZ-class basis states. However, the following derivation is independent of this particular choice and requires only the completeness of the basis. 
Consequently, the maximally mixed state can be written using the completeness relation is therefore
\begin{equation}
\frac{\mathbb{I}_{2^n}}{2^n} = \frac{1}{2^n} |{\rm GHZ}^{+}\rangle \langle{\rm GHZ}^{+}| + \frac{1}{2^n} \sum_{j=2}^{2^n} |\phi_j\rangle \langle\phi_j|.
\label{eq:identity_decomposition}
\end{equation}

Substituting Eq.~(\ref{eq:identity_decomposition}) into Eq.~(17), we obtain
\begin{equation}
W_n(\lambda) = \left( 1-\lambda+\frac{\lambda}{2^n} \right) |{\rm GHZ}^{+}\rangle \langle{\rm GHZ}^{+}| + \frac{\lambda}{2^n} \sum_{j=2}^{2^n} |\phi_j\rangle \langle\phi_j|.
\label{eq:Werner_decomposition}
\end{equation}

The coefficient multiplying the projector $|{\rm GHZ}^{+}\rangle\langle{\rm GHZ}^{+}|$ is precisely the overlap (fidelity) of the noisy ancillary resource with the ideal GHZ state. Hence, the fidelity of a single noisy ancillary GHZ state is
\begin{equation}
F_n(\lambda) = 1-\lambda+\frac{\lambda}{2^n} = 1 - \frac{2^n-1}{2^n}\lambda.
\label{eq:FidelityGHZ}
\end{equation}

For a fixed depolarizing strength $\lambda$, $F_n(\lambda)$ increases with the number of qubits because the maximally mixed component has an overlap $2^{-n}$ with the target GHZ state. Thus, within this depolarizing model, the fidelity exhibits a characteristic system-size dependence that becomes relevant when assessing the robustness of the multipartite ancillary resources. Since the discrimination protocol relies on the ancillary GHZ registers to encode and extract the required computational-pattern and global-phase information, imperfections in these resources directly affect the protocol performance. We therefore next consider the combined effect of noise on the two ancillary registers and its consequent impact on the strict nondestructive success probability.

\subsection{Success probability with two noisy ancillas}


\begin{figure}[t]
\centering
\includegraphics[width=0.55\columnwidth]{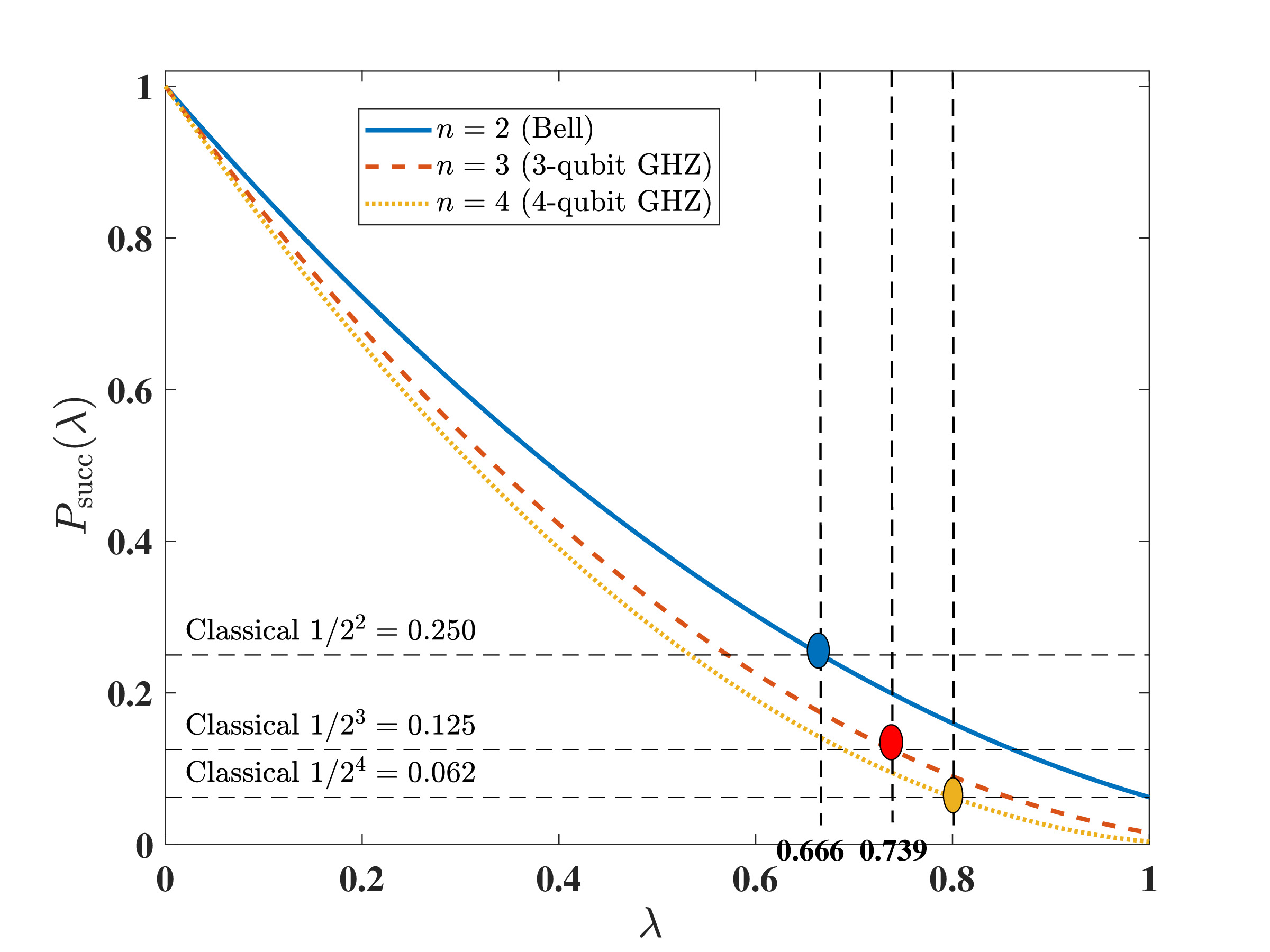}
\caption{Strict nondestructive success probability $P_{\mathrm{succ}}(\lambda)=F_n(\lambda)^2$ for discrimination of $n$-qubit GHZ-class states in the presence of depolarizing noise acting on the ancillary GHZ resources. Results are shown for $n=2$, $3$, and $4$, where $n=2$ corresponds to the Bell-state case, while $n=3$ and $4$ correspond to multipartite GHZ-class states. The dashed horizontal lines indicate the classical nondestructive discrimination probability $P_{\mathrm{cl}}=2^{-n}$. The intersections between the analytical curves and the corresponding classical limits determine the critical depolarizing-noise threshold $\lambda_c$, below which the proposed quantum protocol outperforms the classical benchmark.}
\label{fig:Psucc_lambda_strict_condition}
\end{figure}

The nondestructive discrimination protocol requires two ancillary GHZ states. When both ancillas are ideally prepared in the state $|{\rm GHZ}^+\rangle \equiv \ket{\psi_{0^n}^{+}}$, the protocol succeeds deterministically, as established in the previous sections. We now consider the case in which each ancillary register is described by the depolarized state $W_n(\lambda)$. The imperfections in the ancillary resources then reduce the probability of satisfying the strict nondestructive discrimination criterion.

We define strict nondestructive success as the simultaneous occurrence of two conditions: (i) the ancillary measurement outcomes identify the GHZ-class state according to one of the valid discrimination patterns listed in Table~\ref{tab:3qubit_GHZ}, and (ii) the system qubits remain in their original GHZ-class state. The second condition is essential, since correct identification alone does not constitute nondestructive discrimination if the system state is disturbed.

For the depolarized ancillary state, Eq.~\eqref{eq:WernerGHZ} gives the weight of the desired $|{\rm GHZ}^+\rangle$ component as $F_n(\lambda)$. The remaining weight belongs to the orthogonal noise sector. The discrimination circuit relies on the multipartite correlations of the ideal GHZ ancillas to transfer the parity and phase information to the ancillary registers without disturbing the system. For the strict nondestructive criterion, the contribution from the desired sector of each ancillary resource therefore determines the surviving nondestructive component. Since the two ancillary resources are prepared independently, their contributions factorize. Thus, for ancilla fidelities $F_1$ and $F_2$, the strict nondestructive success probability is

\begin{equation}
P_{\rm succ}=F_1F_2.
\label{eq:GeneralSuccess}
\end{equation}

When both ancillary GHZ states are subjected to the same depolarizing channel, $F_1=F_2=F_n(\lambda)$, and hence

\begin{equation}
P_{\rm succ}(\lambda)
=
\left[
1-\frac{2^n-1}{2^n}\lambda
\right]^2.
\label{eq:SuccessProbability}
\end{equation}

The quadratic dependence in Eq.~\eqref{eq:SuccessProbability} follows directly from the use of two independent ancillary resources. One ancilla is associated with the extraction of the computational-pattern (parity) information, while the other is associated with the extraction of the relative-phase information. Strict nondestructive discrimination requires both resources to provide the required GHZ correlations; therefore, their contributions enter multiplicatively rather than through a single fidelity factor.

To quantify the robustness of the protocol, we compare the above success probability with the classical random-guessing benchmark. An $n$-qubit GHZ-class basis contains $2^n$ mutually orthogonal states, so randomly guessing the unknown label gives

\begin{equation}
P_{\rm cl}=\frac{1}{2^n}.
\label{eq:Classical}
\end{equation}

We therefore identify a quantum advantage relative to this benchmark whenever

\begin{equation}
P_{\rm succ}(\lambda)>P_{\rm cl}.
\end{equation}

Equating the two probabilities gives the corresponding critical depolarizing probability,

\begin{equation}
\lambda_c
=
\frac{2^n}{2^n-1}
\left(
1-\frac{1}{2^{n/2}}
\right),
\label{eq:lambdac}
\end{equation}

such that the protocol remains above the classical benchmark for $\lambda<\lambda_c$.

The dependence of $\lambda_c$ on the number of qubits is noteworthy. Although the fidelity of each ancillary GHZ state decreases with increasing system size for a fixed depolarizing probability, the classical random-guessing benchmark decreases exponentially as $2^{-n}$. As a result, the noise threshold $\lambda_c$ increases with $n$. Thus, relative to the random-guessing benchmark, the protocol retains a quantum advantage over a broader range of depolarizing noise as the number of qubits increases.

Figure~\ref{fig:Psucc_lambda_strict_condition} shows the analytical strict nondestructive success probability as a function of the depolarizing probability for $n=2$, $3$, and $4$. The $n=2$ case corresponds to the bipartite Bell-state limit and is included for comparison, whereas $n=3$ and $n=4$ represent genuinely multipartite GHZ-class systems. In all cases, $P_{\rm succ}$ decreases monotonically with increasing $\lambda$, reflecting the reduction in the fidelity of the ancillary GHZ resources.

The dashed horizontal lines in Fig.~\ref{fig:Psucc_lambda_strict_condition} denote the corresponding classical benchmarks $P_{\rm cl}=2^{-n}$. Their intersections with the analytical curves determine the critical noise thresholds given by Eq.~\eqref{eq:lambdac}. The shift of these intersections toward larger values of $\lambda$ with increasing $n$ illustrates the increasing robustness of the protocol relative to the classical random-guessing benchmark.

These results provide a closed-form characterization of the strict nondestructive performance of the protocol in the presence of depolarizing noise on the ancillary GHZ resources. The depolarized GHZ model therefore provides a simple analytical framework for quantifying how imperfections in the preshared multipartite resources affect the operation of the discrimination protocol.

Throughout this work, successful discrimination is defined according to the strict nondestructive criterion, which requires preservation of the system state after the discrimination protocol. For completeness, Appendix~\ref{appendix:operational_success_probability} discusses an alternative operational definition in which only the ancillary measurement patterns are counted, without explicitly verifying system preservation.


\subsection{Amplitude damping noise on n-qubit GHZ ancillary state}

\begin{figure*}[t]
\centering
\includegraphics[width=0.99\textwidth]{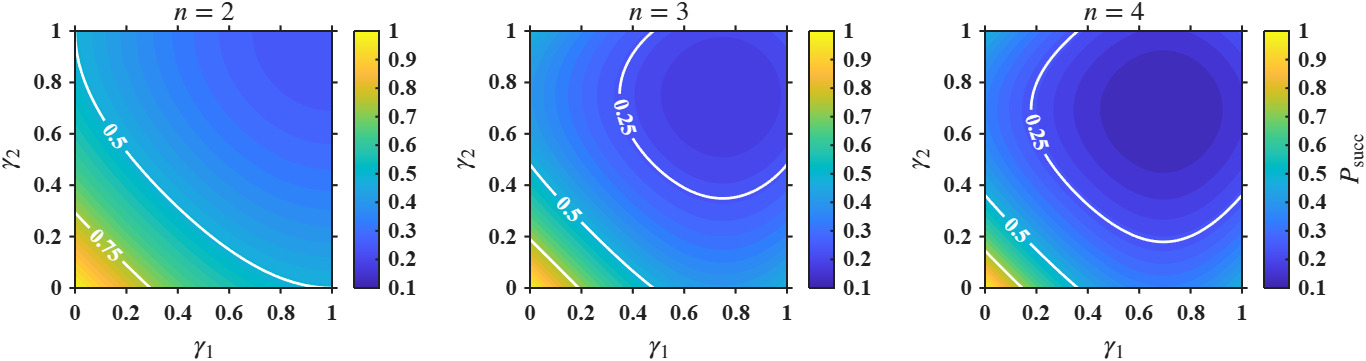}
\caption{
Success probability $P_{\mathrm{succ}}(\gamma_1,\gamma_2)$ for two independently amplitude-damped $n$-qubit GHZ resources as a function of the corresponding damping parameters $\gamma_1$ and $\gamma_2$. The three panels correspond to $n=2$, $3$, and $4$, from left to right, respectively. The $n = 2$ case corresponds to the Bell state and is included for comparison, while $n=3$ and $4$ correspond to the GHZ-class states considered in the present work. The common color scale highlights the progressive reduction in success probability with increasing damping and system
size.}
\label{fig:Psuccess_AD}
\end{figure*}

The preceding subsection considered depolarizing noise because its symmetry-preserving nature allows a complete analytical treatment of the strict nondestructive success probability. In practice, however, ancillary GHZ states may also be affected by dissipative processes, for which amplitude damping provides the standard physical model.

Here, we consider independent amplitude damping acting on each ancillary qubit with damping probability $\gamma$. Unlike depolarizing noise, amplitude damping does not preserve the GHZ symmetry because excitation-loss events preferentially relax the state $|1\rangle$ to $|0\rangle$. As a result, the coherence between the computational components $|0^n\rangle$ and $|1^n\rangle$ is progressively suppressed while population is transferred to computational-basis states outside the two-dimensional GHZ subspace.

The exact amplitude-damped GHZ ancillary state together with its fidelity with the ideal GHZ resource is derived analytically in Appendix~\ref{appendix:Amplitude-Damped GHZ Ancillary Resource}. The resulting fidelity is
\begin{equation}
F_{\rm AD}(\gamma)
=
\frac14
\left[
1
+
(1-\gamma)^n
+
2(1-\gamma)^{n/2}
+
\gamma^n
\right].
\label{Fidelity Aamplitude Damping}
\end{equation}



As shown in Appendix~\ref{appendix:Amplitude-Damped GHZ Ancillary Resource}, the amplitude-damped density matrix of a single $n$-qubit GHZ ancilla, expressed in the GHZ basis, takes the form of Eq.~(\ref{rho_out}), with the population of the ideal ancillary state $|\text{GHZ}^+\rangle\langle\text{GHZ}^+|$ given by the coefficient $C_{++}(\gamma)$. Since the discrimination protocol employs a pair of ancillary $|\text{GHZ}^+\rangle$ states, the protocol succeeds nondestructively precisely when the joint ancillary state remains in the component $|\text{GHZ}^+\,\text{GHZ}^+\rangle\langle \text{GHZ}^+\,\text{GHZ}^+|$, which occurs with probability $C_{++}(\gamma_1)\,C_{++}(\gamma_2)$ for independent damping probabilities $\gamma_1$ and $\gamma_2$ acting on the two resources. The exact strict nondestructive success probability under independent amplitude damping is therefore
\begin{align}
    P_{\rm succ}(\gamma_1,\gamma_2) &= C_{++}(\gamma_1)\,C_{++}(\gamma_2) \nonumber \\
    &= \frac{1}{16}\Big(1 + \gamma_1^n + (1-\gamma_1)^n + 2(1-\gamma_1)^{n/2}\Big)\nonumber\\
   &\times \Big(1 + \gamma_2^n + (1-\gamma_2)^n + 2(1-\gamma_2)^{n/2}\Big).
    \label{eq:SuccessProbAD}
\end{align}



Since the discrimination protocol employs two independent ancillary GHZ resources, the overall quality of the ancillary resource depends on the combined quality of both states. Assuming that the two ancillary GHZ states are prepared independently and experience independent amplitude-damping channels with damping probabilities $\gamma_1$ and $\gamma_2$, their individual fidelities are given by Eq.~(\ref{Fidelity Aamplitude Damping}) as $F_{\rm AD}(\gamma_1)$ and $F_{\rm AD}(\gamma_2)$, respectively. 
For independent amplitude-damping channels, Eq.~(\ref{eq:SuccessProbAD}) can equivalently be expressed in terms of the individual ancillary fidelities as
\begin{equation}
P_{\mathrm{succ}}(\gamma_1,\gamma_2)
=
F_{\mathrm{AD}}(\gamma_1)F_{\mathrm{AD}}(\gamma_2).
\label{eq:JointFidelityAD}
\end{equation}


Figure~\ref{fig:Psuccess_AD} illustrates the dependence of the success probability $P_{\rm succ}(\gamma_1,\gamma_2)$ on the independent damping probabilities $\gamma_1$ and $\gamma_2$ acting on the two ancillary GHZ resources. As expected, unit success probability is attained only in the absence of damping, i.e., $\gamma_1=\gamma_2=0$, and the success probability decreases as either damping probability increases. The degradation becomes progressively more pronounced with increasing number of ancillary qubits. The $n=2$ case is included as a bipartite Bell-state benchmark, while the $n=3$ and $n=4$ cases correspond to the genuinely multipartite GHZ-class resources considered in this work. This increasing degradation indicates the enhanced susceptibility of larger multipartite GHZ states to excitation loss. This behavior arises from the combined effects of the suppression of GHZ-state coherence and the irreversible transfer of population to computational-basis states outside the GHZ subspace due to spontaneous emission. Consequently, amplitude-damping noise produces a more pronounced degradation of the joint ancillary resource and of the protocol's actual success probability, as the system size increases, compared with a symmetry-preserving depolarizing channel.

\subsection{Decoherence Sensing Interpretation}
\label{sec:decoherence_sensing}

The analysis presented above also allows the proposed scheme to be viewed from a different perspective, namely that of quantum decoherence sensing. Rather than regarding the ancillary GHZ states only as resources required for nondestructive state discrimination, they can be treated as calibrated multipartite probes of the noise affecting the quantum network. A known GHZ state is prepared as a reference resource and subsequently subjected to the same physical conditions as the distributed ancillary resources. Comparing the resulting state with the ideal reference then provides a quantitative measure of the degradation caused by the surrounding noise \cite{rossi2015entangled}.

This interpretation is particularly relevant for GHZ states because their multipartite coherence is sensitive to local decoherence \cite{carvalho2004decoherence}. In the depolarizing model considered above, the fidelity with respect to the ideal $n$-qubit GHZ state, given by Eq.~(\ref{eq:FidelityGHZ}), directly quantifies the weight of the desired GHZ component in the noisy resource. For independent amplitude damping, the corresponding fidelity in Eq.~(\ref{Fidelity Aamplitude Damping}) quantifies the degradation resulting from excitation loss. Thus, once the relevant noise model has been characterized, the measured GHZ-state fidelity can be related to the corresponding decoherence parameter.

The use of two ancillary GHZ resources provides an additional feature in this context. In the depolarizing model considered above, both resources are taken to experience the same noise strength~$\lambda$, whereas in the amplitude-damping analysis the two ancillary states may be characterized independently by damping parameters~$\gamma_1$ and~$\gamma_2$. Their individual fidelities quantify the degradation of each resource, while the corresponding joint fidelity characterizes the quality of the complete multipartite ancillary state used by the discrimination protocol. In this sense, the protocol provides an operational setting in which the degradation of known entangled resources can be monitored through experimentally accessible quantities.

This viewpoint motivates the interpretation of the proposed scheme as a quantum decoherence meter. In such an interpretation, a known multipartite entangled state acts as a calibrated probe, the ancillary GHZ resources are exposed to the noise present in the surrounding environment, and their measured fidelity provides a quantitative indication of the corresponding decoherence strength. The dependence on the number of qubits is also relevant: larger GHZ states generally exhibit a stronger response to local decoherence \cite{carvalho2004decoherence}. This may make larger multipartite probes useful for detecting weak noise, although the increased sensitivity is accompanied by the greater fragility of the GHZ resource itself \cite{huelga1997improvement}.

We stress that this interpretation should not be confused with a fully optimized quantum-sensing protocol. The present work establishes analytical relations between the degradation of the GHZ resources and the parameters of the specific noise models considered here, but it does not optimize the estimation procedure, calculate the quantum Fisher information, or determine the ultimate precision bounds of the probe \cite{degen2017quantum,nolan2017quantum}. Accordingly, we use the term decoherence sensing to describe the operational interpretation of the present results. A detailed investigation of parameter-estimation precision, Fisher-information scaling, and optimal sensing strategies based on the proposed multipartite resources is left for future work.

\section{Experimental feasibility}
\label{sec:experimental_feasibility}

\begin{figure}[t]
\centering
\includegraphics[width=0.48\textwidth]{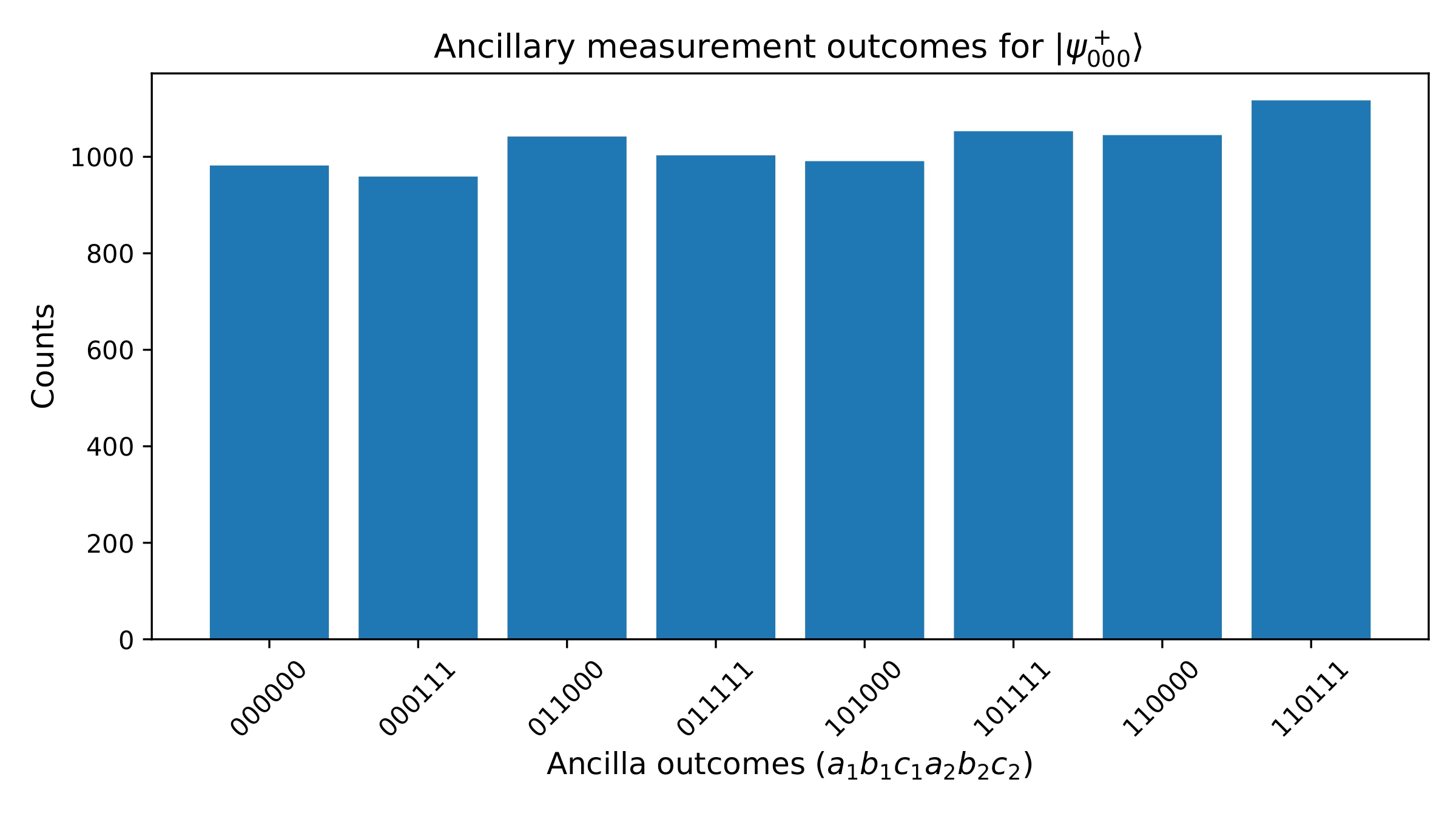}
\hfill
\includegraphics[width=0.48\textwidth]{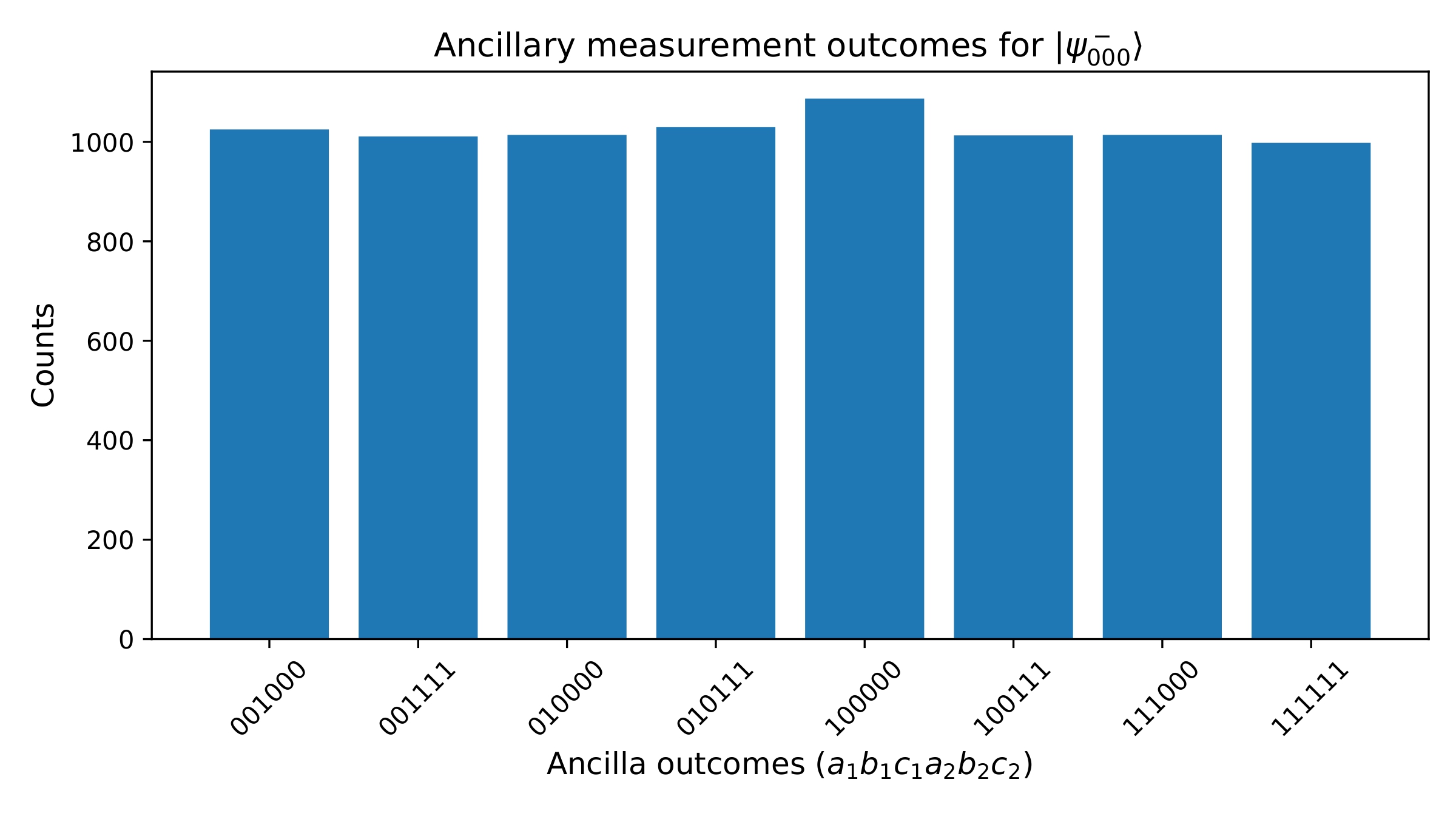}

\caption{
Histograms of ancillary measurement outcomes obtained from Qiskit simulations of the nondestructive discrimination protocol for two representative three-qubit GHZ-class states listed in Table~\ref{tab:3qubit_GHZ}. The simulation was performed with $8192$ measurement shots. The horizontal axis shows the ancillary outcomes $(a_1 b_1 c_1 a_2 b_2 c_2)$, while the vertical axis denotes the corresponding measurement counts.
(a) Results for $\ket{\psi^+_{000}}=\frac{1}{\sqrt{2}}(\ket{000}+\ket{111})$.  
(b) Results for $\ket{\psi^-_{000}}=\frac{1}{\sqrt{2}}(\ket{000}-\ket{111})$.
In each case, only the eight theoretically allowed bit strings predicted in Table~\ref{tab:3qubit_GHZ} occur with nonzero probability, confirming correct discrimination. The complementary outcome patterns between panels (a) and (b) demonstrate the phase sensitivity of the protocol. The system qubits remain unmeasured, and the post-interaction fidelity with the initial GHZ state remains approximately unity, verifying nondestructive operation. Small variations in bar heights arise from finite sampling statistics.
}

\label{fig:histograms}
\end{figure}

The proposed multipartite protocol does not require any operations beyond those already available on current quantum hardware. Its implementation relies only on parallel local system–ancilla interactions together with GHZ-state preparation, making the protocol compatible with existing trapped-ion and superconducting architectures.

A proof-of-principle implementation of the three-qubit GHZ protocol requires nine qubits in total, which is within the qubit range of current trapped-ion processors. The system GHZ state can be verified to remain unchanged through conditional state tomography based on the ancillary measurement outcomes Ref.~\cite{NondestructivePRA2025}. In contrast to the experimentally demonstrated bipartite Bell-state protocol, the multipartite scheme requires only parallelization of identical system–ancilla interactions without introducing additional nonlocal resources. We therefore expect that near-term quantum processors with sufficient qubit number and coherence time should be capable of implementing the present protocol.

To demonstrate the proposed protocol, we implemented the three-qubit protocol using the Qiskit software framework \cite{javadi2024quantum} under ideal unitary dynamics. The circuit directly follows the theoretical construction, including GHZ preparation, local CNOT interactions, Hadamard operations on the first ancilla, and projective measurements on ancillary qubits only. The explicit circuit is provided in Appendix. \ref{app:circuit}.

Representative simulation results are shown in Fig.~\ref{fig:histograms} for two GHZ-class states listed in Table~\ref{tab:3qubit_GHZ}. Figure~\ref{fig:histograms}(a) corresponds to the input state $\ket{\psi^+_{000}}=\frac{1}{\sqrt2}(\ket{000}+\ket{111})$, while Fig.~\ref{fig:histograms}(b) corresponds to $\ket{\psi^-_{000}}=\frac{1}{\sqrt2}(\ket{000}-\ket{111})$. In both cases, the simulation yields exactly eight ancillary measurement outcomes with nonzero probability, in precise agreement with the theoretical predictions summarized in Table~\ref{tab:3qubit_GHZ}. All remaining $2^6-8$ computational-basis outcomes are suppressed up to numerical precision.

Importantly, the complementary bit-string patterns observed between panels (a) and (b) demonstrate that the protocol correctly distinguishes the relative phase of the GHZ state. Furthermore, the reduced density matrix of the system qubits after completion of the protocol has unit fidelity with the initial GHZ state in both cases, confirming that the discrimination process is nondestructive.

These numerical results provide explicit verification of the parity–phase mapping mechanism underlying the protocol and demonstrate its correctness under ideal unitary dynamics. All remaining GHZ-class states listed in Table~\ref{tab:3qubit_GHZ} were similarly verified and exhibit identical agreement with the analytical predictions. Beyond reproducing the analytical predictions, the simulations verify that the complete distributed circuit implements the intended parity–phase mapping while preserving the system state throughout the protocol.

The circuit structure generalizes straightforwardly to the four-qubit and arbitrary $n$-qubit cases. The implementation requires two ancillary $n$-qubit GHZ states and parallel controlled-NOT operations between corresponding system and ancillary qubits. No additional multi-qubit interactions beyond nearest-pair CNOT gates are required.

The total number of qubits required scales linearly as $3n$ ($n$ system qubits and $2n$ ancillary qubits), while the circuit depth increases only by a constant number of additional gate layers compared to the Bell-state protocol. Therefore, the protocol preserves a simple distributed architecture in which scalability is determined primarily by available qubit number and coherence time rather than by increasing logical circuit complexity.

\section{Conclusion}
\label{sec:conclusion}
In this work we have presented a generalized and scalable protocol for the nondestructive discrimination of distributed GHZ-class states assisted by ancillary GHZ resources. We derived explicit discrimination rules and measurement signatures for three- and four-qubit GHZ-class states and showed that the protocol naturally generalizes to arbitrary system size $n$. Importantly, the scheme requires local unitary operations together with projective measurements on ancillary qubits, while the system state itself remains preserved. This provides a concrete operational framework for extracting multipartite entanglement information without disturbing the encoded quantum state.

The present work establishes a unified distributed framework for strict nondestructive discrimination of GHZ-class states of arbitrary size, in which the global parity--phase information encoded in multipartite entanglement is coherently transferred to two ancillary registers while the system state remains unchanged. From a scalability perspective, the number of ancillary qubits scales linearly with system size, while the ancillary measurement space grows exponentially. Nevertheless, the symmetry of GHZ states restricts the physically accessible outcomes to only $2^{n}$ correlated bit patterns for each GHZ-class state, enabling efficient and unambiguous discrimination for larger systems. The protocol therefore provides a practical framework for accessing global multipartite correlations using preshared entangled ancillary resources.

We further investigated realistic scenarios in which the ancillary GHZ resources are affected by noise, considering both depolarizing and amplitude-damping channels acting on the two ancillary registers. For the depolarizing model, we obtained a closed-form expression for the success probability and determined the critical noise thresholds beyond which the quantum protocol no longer outperforms classical nondestructive strategies. For amplitude damping, which is a non-unital dissipative process that does not preserve the GHZ symmetry, we derived the exact damped ancillary state and showed that the strict nondestructive success probability factorizes into the product of the contributions from the two ancillary resources. Although the amplitude-damped state retains residual coherence between the $|\mathrm{GHZ}^{+}\rangle$ and $|\mathrm{GHZ}^{-}\rangle$ sectors, the projective measurement selects the $|\mathrm{GHZ}^{+}\rangle$ component relevant for successful discrimination. The success probability decreases as the damping increases, with stronger degradation for larger ancillary GHZ-class states. This shows that larger GHZ resources are more sensitive to amplitude-damping noise.



Finally, the analysis also suggests a broader interpretation of the proposed scheme as a form of quantum decoherence sensing. The ancillary GHZ resources can be viewed as calibrated multipartite probes whose degradation provides information about the noise experienced by the distributed quantum system. Since the ideal GHZ state is known and the corresponding noisy-state fidelities are explicitly related to the parameters of the depolarizing and amplitude-damping channels considered here, the measured degradation of the ancillary resources can be used to characterize the effective decoherence strength. The use of two ancillary resources further allows the degradation of the resources participating in the parity and phase extraction processes to be monitored within the same operational framework. In this sense, the discrimination protocol can also serve as a primitive for a quantum decoherence meter, in which multipartite entangled states act as calibrated probes of the surrounding quantum environment. The present work does not, however, address optimal parameter estimation or metrological sensitivity; these questions, including Fisher-information analysis and optimization of the probe size and measurement strategy, remain open for future investigation.

Several directions therefore remain open for future investigation. Because the protocol preserves the shared GHZ resource after discrimination, it may serve as a useful primitive for distributed quantum communication and multipartite cryptographic protocols that require repeated verification of shared entangled resources. The nondestructive extraction of parity--phase information may also be useful for diagnosing noise processes, monitoring entanglement in distributed networks, and developing nondestructive error-syndrome extraction schemes. The decoherence-sensing interpretation developed above also motivates a systematic study of the precision with which environmental noise parameters can be estimated using multipartite GHZ probes. It would be particularly interesting to investigate the dependence of the sensing performance on system size, ancilla preparation quality, and the structure of the underlying noise channel. More broadly, alternative quantum resources, including temporal correlations associated with indefinite causal order, could be explored as possible substitutes or complements to preshared entangled ancillas in nondestructive discrimination protocols. These directions may further broaden the role of strict nondestructive discrimination as a practical building block for distributed quantum information processing and quantum sensing.

\begin{acknowledgments}
The authors gratefully acknowledge Siksha 'O' Anusandhan (SOA) Deemed to be University, Bhubaneswar, Odisha, India, for providing the institutional support and necessary facilities for carrying out this research.


\end{acknowledgments}



\appendix

\section{Operational success probability without system verification}
\label{appendix:operational_success_probability}

The main text adopts a strict definition of nondestructive discrimination, in which successful identification of the GHZ-class state must be accompanied by exact preservation of the system state. For completeness, we briefly distinguish this criterion from a weaker operational definition in which only the ancillary measurement outcomes are used to infer the GHZ-class label, without explicitly verifying the state of the system after the protocol.

For the depolarized ancillary resource, Eq.~\eqref{eq:FidelityGHZ} may be written as \begin{equation} 
W_n(\lambda) = F_n(\lambda) |{\rm GHZ}^{+}\rangle\langle{\rm GHZ}^{+}| + \left[1-F_n(\lambda)\right]\rho_{\rm noise},
\end{equation} 
where $\rho_{\rm noise}$ is supported on the subspace orthogonal to $|{\rm GHZ}^{+}\rangle$. Thus, $F_n(\lambda)$ gives the weight of the desired GHZ component in a single ancillary resource, while the remaining weight belongs to the orthogonal noise sector.

With two independently prepared ancillary resources, the joint state can consequently be regarded as a statistical mixture of four components,
\begin{align} &|{\rm GHZ}^{+}\rangle\langle{\rm GHZ}^{+}| \otimes |{\rm GHZ}^{+}\rangle\langle{\rm GHZ}^{+}|, \nonumber\\
&|{\rm GHZ}^{+}\rangle\langle{\rm GHZ}^{+}| \otimes \rho_{\rm noise}, \nonumber\\ 
&\rho_{\rm noise}\otimes |{\rm GHZ}^{+}\rangle\langle{\rm GHZ}^{+}|, \nonumber\\
&\rho_{\rm noise}\otimes\rho_{\rm noise},
\end{align} 
with corresponding weights $F_n^2$, $F_n(1-F_n)$, $F_n(1-F_n)$, and $(1-F_n)^2$.

The first component reproduces the ideal protocol and therefore gives the deterministic nondestructive discrimination result. The other three components correspond to cases in which at least one of the ancillary resources has left the desired GHZ sector. Their contribution to the ancillary measurement statistics depends on the detailed structure of the noise sector and on the action of the discrimination circuit. In particular, the orthogonal component $\rho_{\rm noise}$ cannot, in general, be identified with a uniformly random distribution over the valid discrimination outcomes.

Consequently, an ancilla-only operational success probability cannot be obtained solely from the single-ancilla fidelity $F_n(\lambda)$ without explicitly propagating the corresponding noisy density matrices through the discrimination circuit. Such a quantity would require evaluating the probability of obtaining the correct discrimination patterns while not imposing the additional condition that the system state remain unchanged.

This distinction is important for interpreting the result of the present work. The quantity $P_{\rm succ}=F_n(\lambda)^2$ derived in the main text is specifically the strict nondestructive success probability. It counts only those events for which the ancillary resources provide the required parity--phase correlations and the system state is preserved. A weaker ancilla-only success probability may receive additional contributions from noisy sectors, but these contributions cannot be inferred from the ancillary fidelity alone.

We therefore do not use an ancilla-only operational success probability in the main analysis. The strict criterion is retained throughout the paper because it directly captures the defining requirement of nondestructive discrimination: the identified quantum state must remain available for subsequent use.

\section{Amplitude-Damped GHZ Ancillary Resource}
\label{appendix:Amplitude-Damped GHZ Ancillary Resource}

In this appendix, we derive the exact density matrix of an $n$-qubit GHZ ancillary resource subjected to independent amplitude-damping noise acting on each ancillary qubit. The resulting state is then used to evaluate the fidelity of the noisy ancillary resource with respect to the ideal GHZ state employed in the discrimination protocol. This analysis complements the depolarizing-noise model considered in the main text by providing an exact characterization of a physically relevant, non-unital noise channel acting on the ancillary entangled resource.

The ideal ancillary GHZ state is
\begin{equation}
|{\rm GHZ}^{+}\rangle
=
\frac{1}{\sqrt{2}} \left( |0^{\otimes n}\rangle + |1^{\otimes n}\rangle \right),
\label{eq:GHZstate}
\end{equation}

The standard $n$-qubit GHZ basis consists of $2^n$ orthonormal states indexed by a bitstring $k = (k_1 k_2 \dots k_n) \in \{0, 1\}^n$, where $k_1$ sets the relative phase and $k_2 \dots k_n$ fix the target bit pattern:
\begin{equation}
|\text{GHZ}_k^\pm\rangle = \frac{1}{\sqrt{2}} \left( |0\rangle|k_2 \dots k_n\rangle \pm |1\rangle|\bar{k}_2 \dots \bar{k}_n\rangle \right).
\end{equation}

Restricting to the two-dimensional subspace spanned by $\{|0^{\otimes n}\rangle, |1^{\otimes n}\rangle\}$, the relevant basis states reduce to
\begin{align}
|\text{GHZ}^+\rangle &= \frac{1}{\sqrt{2}} \left( |0^{\otimes n}\rangle + |1^{\otimes n}\rangle \right), \\
|\text{GHZ}^-\rangle &= \frac{1}{\sqrt{2}} \left( |0^{\otimes n}\rangle - |1^{\otimes n}\rangle \right).
\end{align}

For every remaining bitstring $\nu \notin \{0^{\otimes n}, 1^{\otimes n}\}$, the paired computational-basis states $|\nu\rangle$ and $|\bar{\nu}\rangle$ generate orthogonal GHZ-like basis pairs
\begin{equation}
|\Psi_\nu^\pm\rangle = \frac{1}{\sqrt{2}} \left( |\nu\rangle \pm |\bar{\nu}\rangle \right).
\end{equation}

The action of the amplitude-damping channel on the $n$-qubit ancillary state $\rho_{\rm GHZ} = |{\rm GHZ}^{+}\rangle\langle{\rm GHZ}^{+}|$ is given by the Kraus decomposition
\begin{equation}
\rho_{\rm AD}
=
\sum_{\mu}
K_\mu
\rho_{\rm GHZ}
K_\mu^\dagger,
\label{eq:rhoAD}
\end{equation}
where the index $\mu$ labels every combination of single-qubit Kraus operators,
\begin{equation}
K_\mu
=
E_{i_1}\otimes E_{i_2}\otimes\cdots\otimes E_{i_n},
\qquad
i_k\in\{0,1\},
\end{equation}
with
\begin{align}
E_0
&=
|0\rangle\langle0|
+
\sqrt{1-\gamma}\,
|1\rangle\langle1|,
\\
E_1
&=
\sqrt{\gamma}\,
|0\rangle\langle1|,
\end{align}
and $\gamma$ denotes the amplitude-damping probability.

Applying the global Kraus operators $\{K_\mu\}$ to $\rho_{\rm GHZ}$, only the no-jump operator $E_0^{\otimes n}$ acts nontrivially on both components $|0^{\otimes n}\rangle$ and $|1^{\otimes n}\rangle$ simultaneously, thereby preserving their mutual coherence. Any global Kraus operator containing at least one occurrence of $E_1$ annihilates the $|0^{\otimes n}\rangle$ component while mapping $|1^{\otimes n}\rangle$ onto a different computational-basis state; the coherence between $|0^{\otimes n}\rangle$ and $|1^{\otimes n}\rangle$ is therefore completely lost along every such quantum-jump trajectory. The resulting $n$-qubit density matrix, expressed in the GHZ basis, is

\begin{align}\label{rho_out}
\rho_{\text{AD}}(\gamma) &= C_{++}(\gamma) |\text{GHZ}^+\rangle\langle\text{GHZ}^+| + C_{--}(\gamma) |\text{GHZ}^-\rangle\langle\text{GHZ}^-|\nonumber \\
&+ C_{+-}(\gamma) \left( |\text{GHZ}^+\rangle\langle\text{GHZ}^-| + |\text{GHZ}^-\rangle\langle\text{GHZ}^+| \right) \nonumber \\
&+ \frac{1}{4} \sum_{\nu \neq 0^{\otimes n}, 1^{\otimes n}} \gamma^{n-|\nu|} (1-\gamma)^{|\nu|} \big( |\Psi_\nu^+\rangle\langle\Psi_\nu^+| + |\Psi_\nu^-\rangle\langle\Psi_\nu^-| \nonumber\\
& + |\Psi_\nu^+\rangle\langle\Psi_\nu^-| + |\Psi_\nu^-\rangle\langle\Psi_\nu^+| \big),
\end{align}


with coefficients
\begin{align}
C_{++}(\gamma) &= \frac{1}{4} \left( 1 + \gamma^n + (1-\gamma)^n + 2(1-\gamma)^{n/2} \right), \\
C_{--}(\gamma) &= \frac{1}{4} \left( 1 + \gamma^n + (1-\gamma)^n - 2(1-\gamma)^{n/2} \right), \\
C_{+-}(\gamma) &= \frac{1}{4} \left( 1 + \gamma^n - (1-\gamma)^n \right).
\end{align}

Since the discrimination protocol employs the ancillary GHZ state as the nonlocal resource that carries the computational-pattern and phase information, the fidelity with respect to the ideal GHZ state provides a natural quantitative measure of resource degradation under amplitude damping,
\begin{equation}
F_{\rm AD}
=
\langle{\rm GHZ}^{+}|
\rho_{\rm AD}
|{\rm GHZ}^{+}\rangle.
\end{equation}
From Eq.~(\ref{rho_out}), only the populations of $|0^{\otimes n}\rangle$ and $|1^{\otimes n}\rangle$, together with their surviving coherence, contribute to this overlap, giving
\begin{equation}
F_{\rm AD}(\gamma)
=
\frac14
\left[
1
+
(1-\gamma)^n
+
2(1-\gamma)^{n/2}
+
\gamma^n
\right].
\label{eq:FAD}
\end{equation}

Equation~(\ref{eq:FAD}) depends explicitly on both the damping probability $\gamma$ and the number of ancillary qubits $n$, and thus quantifies how the quality of the multipartite GHZ resource deteriorates with increasing system size. It satisfies the expected limiting behavior,
\begin{equation}
F_{\rm AD}(0)=1,
\qquad
F_{\rm AD}(1)=\frac12,
\end{equation}
corresponding, respectively, to the absence of amplitude damping and to complete decay.

Since our nondestructive discrimination protocol employs a pair of $|{\rm GHZ}^{+}\rangle$ states as ancillary resources, the output state of the pair under independent amplitude damping follows directly from Eq.~(\ref{rho_out}):
\begin{equation}
    \rho_{AD}=\rho_{AD}(\gamma_1)\otimes \rho_{AD}(\gamma_2).
\end{equation}
The state $|{\rm GHZ}^{+}\,{\rm GHZ}^{+}\rangle\langle {\rm GHZ}^{+}\,{\rm GHZ}^{+}|$ occurs in this mixture with probability $C_{++}(\gamma_1)\,C_{++}(\gamma_2)$, and it is precisely this component that enables successful nondestructive discrimination of the GHZ-class states. The success probability of the protocol is therefore
\begin{align}
    & P_{succ}(\gamma_1,\gamma_2) = C_{++}(\gamma_1)\,C_{++}(\gamma_2) \nonumber\\
   &= \frac{1}{16}\Big(1 + \gamma_1^n + (1-\gamma_1)^n + 2(1-\gamma_1)^{n/2}\Big)\nonumber\\
   & \quad \times \Big(1 + \gamma_2^n + (1-\gamma_2)^n + 2(1-\gamma_2)^{n/2}\Big).
\end{align}

For weak amplitude damping, $\gamma\ll1$, Eq.~(\ref{eq:FAD}) gives
\begin{equation}
F_{\rm AD}(\gamma)
=
1-\frac{n}{2}\gamma+O(\gamma^2).
\label{eq:FAD_weak}
\end{equation}
Thus, to leading order, the degradation of the GHZ-resource fidelity grows
linearly with the number of ancillary qubits, reflecting the increasing
sensitivity of multipartite coherence to independent local amplitude damping.

\section{Three-Qubit Quantum Circuit Implementation}
\label{app:circuit}

\begin{figure}[t]
\centering
\includegraphics[width=0.95\textwidth]{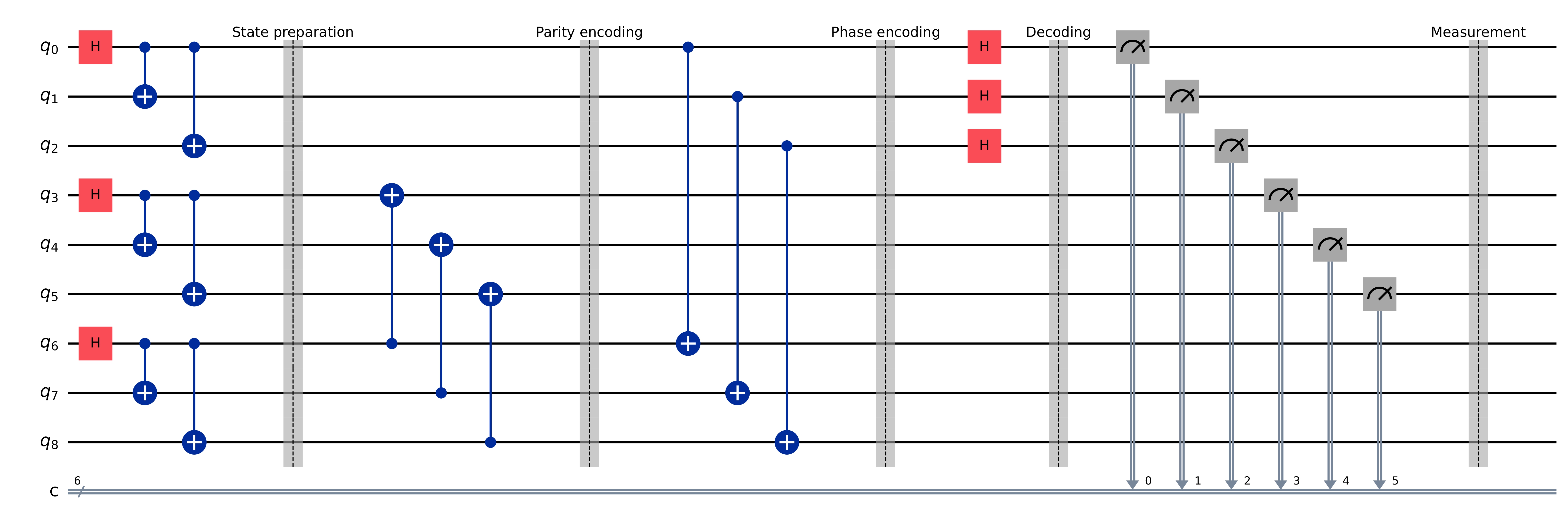}
\caption{
Quantum circuit implementation of the ancilla-assisted nondestructive discrimination protocol for three-qubit GHZ-class states. The circuit includes explicit preparation of the system and ancillary GHZ states using standard Hadamard–CNOT sequences, followed by the parity-encoding and phase-encoding operations described in Sec.~\ref{sec:protocol}. Barriers indicate different logical stages of the protocol. The qubit ordering is $(q_0,\ldots,q_8)\equiv(a_1,b_1,c_1,a_2,b_2,c_2,A,B,C)$. Within the discrimination protocol, the first layer of CNOT gates transfers parity information from the system qubits to the second ancillary register, while the second layer correlates the system with the first ancillary register, enabling the global phase information to be encoded in the first ancillary register. Local Hadamard gates on $(a_1,b_1,c_1)$ convert phase correlations into measurable parity outcomes. Only the ancillary qubits are measured, leaving the system GHZ state unchanged.}
\label{fig:qiskit_circuit}
\end{figure}

Here, we present the explicit quantum circuit implementation of the three-qubit ancilla-assisted nondestructive discrimination protocol used in our numerical simulations. The circuit was implemented using Qiskit \cite{javadi2024quantum} and directly realizes the sequence of unitary operations described in Sec.~\ref{sec:protocol}. For completeness, the circuit also includes explicit preparation of the system and ancillary GHZ states using standard Hadamard and CNOT gate sequences. In the theoretical protocol described in Sec.~\ref{sec:protocol}, these GHZ states are assumed to be preshared among the parties, whereas in the numerical implementation they are prepared explicitly within the circuit. As shown in Fig.~\ref{fig:qiskit_circuit}, the circuit uses nine qubits:
three system qubits and two ancillary GHZ registers.

The qubit ordering in the circuit is
\[
(q_0,q_1,q_2,q_3,q_4,q_5,q_6,q_7,q_8)
\equiv
(a_1,b_1,c_1,a_2,b_2,c_2,A,B,C).
\]
Here $(A,B,C)$ denote the system qubits initially sharing an unknown GHZ-class state, while $(a_1,b_1,c_1)$ and $(a_2,b_2,c_2)$ denote the two ancillary GHZ states preshared among Alice, Bob, and Charlie.

The protocol proceeds in three stages. First, CNOT operations
\[
A \rightarrow a_2, \quad
B \rightarrow b_2, \quad
C \rightarrow c_2
\]
coherently transfer parity information from the system to the second ancillary register. Second, CNOT operations
\[
a_1 \rightarrow A, \quad
b_1 \rightarrow B, \quad
c_1 \rightarrow C
\]
encode phase information through interactions with the first ancillary register. Third, local Hadamard gates are applied to $(a_1,b_1,c_1)$ to convert phase correlations into parity information that can be accessed by measurement.

Finally, projective measurements are performed only on the ancillary qubits $(a_1,b_1,c_1,a_2,b_2,c_2)$, while the system qubits remain unmeasured throughout the protocol. The measurement results are stored in classical bits $(c_0,\ldots,c_5)$ and form the six-bit string used to identify the GHZ-class state. Consequently, the measurement outcomes uniquely determine the GHZ-class state while leaving the system state unchanged, thereby realizing nondestructive discrimination. Numerical simulations confirm that the post-interaction system state retains unit fidelity with the initial GHZ state, providing an explicit circuit-level verification of the nondestructive character of the protocol.

\bibliographystyle{apsrev4-2}
\bibliography{sn-bibliography}

@article{wootters1982single,
  title={A single quantum cannot be cloned},
  author={Wootters, William K and Zurek, Wojciech H},
  journal={Nature},
  volume={299},
  number={5886},
  pages={802--803},
  year={1982},
  publisher={Nature Publishing Group UK London}
}

@article{lindblad1999general,
  title={A general no-cloning theorem},
  author={Lindblad, G{\"o}ran},
  journal={Letters in Mathematical Physics},
  volume={47},
  number={2},
  pages={189--196},
  year={1999},
  publisher={Springer}
}

@article{kalev2008no,
  title={No-broadcasting theorem and its classical counterpart},
  author={Kalev, Amir and Hen, Itay},
  journal={Physical review letters},
  volume={100},
  number={21},
  pages={210502},
  year={2008},
  publisher={APS}
}

@article{bennett1999quantum,
  title={Quantum nonlocality without entanglement},
  author={Bennett, Charles H and DiVincenzo, David P and Fuchs, Christopher A and Mor, Tal and Rains, Eric and Shor, Peter W and Smolin, John A and Wootters, William K},
  journal={Physical Review A},
  volume={59},
  number={2},
  pages={1070},
  year={1999},
  publisher={APS}
}

@article{popescu1994causality,
  title={Causality constraints on nonlocal quantum measurements},
  author={Popescu, Sandu and Vaidman, Lev},
  journal={Physical Review A},
  volume={49},
  number={6},
  pages={4331},
  year={1994},
  publisher={APS}
}

@article{cavalcanti2011quantum,
  title={Quantum networks reveal quantum nonlocality},
  author={Cavalcanti, Daniel and Almeida, Mafalda L and Scarani, Valerio and Acin, Antonio},
  journal={Nature communications},
  volume={2},
  number={1},
  pages={184},
  year={2011},
  publisher={Nature Publishing Group UK London}
}

@article{popescu1992generic,
  title={Generic quantum nonlocality},
  author={Popescu, Sandu and Rohrlich, Daniel},
  journal={Physics Letters A},
  volume={166},
  number={5-6},
  pages={293--297},
  year={1992},
  publisher={Elsevier}
}

@article{popescu1994quantum,
  title={Quantum nonlocality as an axiom},
  author={Popescu, Sandu and Rohrlich, Daniel},
  journal={Foundations of Physics},
  volume={24},
  number={3},
  pages={379--385},
  year={1994},
  publisher={Springer}
}

@article{bhattacharya2020nonlocality,
  title={Nonlocality without entanglement: Quantum theory and beyond},
  author={Bhattacharya, Some Sankar and Saha, Sutapa and Guha, Tamal and Banik, Manik},
  journal={Physical Review Research},
  volume={2},
  number={1},
  pages={012068},
  year={2020},
  publisher={APS}
}

@article{bennett2014quantum,
  title={Quantum cryptography: Public key distribution and coin tossing},
  author={Bennett, Charles H and Brassard, Gilles},
  journal={Theoretical computer science},
  volume={560},
  pages={7--11},
  year={2014},
  publisher={Elsevier}
}

@article{pirandola2020advances,
  title={Advances in quantum cryptography},
  author={Pirandola, Stefano and Andersen, Ulrik L and Banchi, Leonardo and Berta, Mario and Bunandar, Darius and Colbeck, Roger and Englund, Dirk and Gehring, Tobias and Lupo, Cosmo and Ottaviani, Carlo and others},
  journal={Advances in optics and photonics},
  volume={12},
  number={4},
  pages={1012--1236},
  year={2020},
  publisher={Optical Society of America}
}

@article{xu2020secure,
  title={Secure quantum key distribution with realistic devices},
  author={Xu, Feihu and Ma, Xiongfeng and Zhang, Qiang and Lo, Hoi-Kwong and Pan, Jian-Wei},
  journal={Reviews of modern physics},
  volume={92},
  number={2},
  pages={025002},
  year={2020},
  publisher={APS}
}

@article{bouwmeester1997experimental,
  title={Experimental quantum teleportation},
  author={Bouwmeester, Dik and Pan, Jian-Wei and Mattle, Klaus and Eibl, Manfred and Weinfurter, Harald and Zeilinger, Anton},
  journal={Nature},
  volume={390},
  number={6660},
  pages={575--579},
  year={1997},
  publisher={Nature Publishing Group UK London}
}

@article{jin2015highly,
  title={Highly efficient entanglement swapping and teleportation at telecom wavelength},
  author={Jin, Rui-Bo and Takeoka, Masahiro and Takagi, Utako and Shimizu, Ryosuke and Sasaki, Masahide},
  journal={Scientific reports},
  volume={5},
  number={1},
  pages={9333},
  year={2015},
  publisher={Nature Publishing Group UK London}
}

@article{gupta2007general,
  title={General circuits for indirecting and distributing measurement in quantum computation},
  author={Gupta, M and Pathak, A and Srikanth, R and Panigrahi, Prasanta K.},
  journal={International Journal of Quantum Information},
  volume={5},
  number={04},
  pages={627--640},
  year={2007},
  publisher={World Scientific}
}

@inproceedings{panigrahi2006circuits,
  title={Circuits for distributing quantum measurement},
  author={Panigrahi, Prasanta K. and Gupta, M and Pathak, A and Srikanth, R},
  booktitle={AIP Conference Proceedings},
  year={2006},
  organization={American Institute of Physics}
}

@article{wang2013nondestructive,
  title={Nondestructive greenberger-horne-zeilinger-state analyzer},
  author={Wang, Xin-Wen and Zhang, Deng-Yu and Tang, Shi-Qing and Xie, Li-Jun},
  journal={Quantum information processing},
  volume={12},
  pages={1065--1075},
  year={2013},
  publisher={Springer}
}

@article{satyajit2018nondestructive,
  title={Nondestructive discrimination of a new family of highly entangled states in IBM quantum computer},
  author={Satyajit, Saipriya and Srinivasan, Karthik and Behera, Bikash K and Panigrahi, Prasanta K},
  journal={Quantum Information Processing},
  volume={17},
  number={9},
  pages={212},
  year={2018},
  publisher={Springer}
}

@article{NondestructivePRA2025,
  title = {Nondestructive discrimination of Bell states between distant parties},
  author = {Bilash, Bohdan and Lim, Youngrong and Kwon, Hyukjoon and Kim, Yosep and Lim, Hyang-Tag and Song, Wooyeong and Kim, Yong-Su},
  journal = {Phys. Rev. A},
  volume = {110},
  issue = {4},
  pages = {042407},
  numpages = {7},
  year = {2024},
  month = {Oct},
  publisher = {American Physical Society},
}

@article{sisodia2017experimental,
  title={Experimental realization of nondestructive discrimination of Bell states using a five-qubit quantum computer},
  author={Sisodia, Mitali and Shukla, Abhishek and Pathak, Anirban},
  journal={Physics Letters A},
  volume={381},
  number={46},
  pages={3860--3874},
  year={2017},
  publisher={Elsevier}
}

@article{ghosh2018automated,
  title={Automated error correction in IBM quantum computer and explicit generalization},
  author={Ghosh, Debjit and Agarwal, Pratik and Pandey, Pratyush and Behera, Bikash K and Panigrahi, Prasanta K.},
  journal={Quantum Information Processing},
  volume={17},
  number={6},
  pages={153},
  year={2018},
  publisher={Springer}
}

@article{thatte2025quantum,
  title={Quantum dialogue through non-destructive discrimination of cluster state},
  author={Thatte, Mandar and Banerjee, Shreya and Panigrahi, Prasanta K.},
  journal={Journal of Physics A: Mathematical and Theoretical},
  volume={58},
  number={50},
  pages={505303},
  year={2025},
  publisher={IOP Publishing}
}

@article{jain2009secure,
  title={Secure quantum conversation through non-destructive discrimination of highly entangled multipartite states},
  author={Jain, Sakshi and Muralidharan, Sreraman and Panigrahi, Prasanta K.},
  journal={EPL (Europhysics Letters)},
  volume={87},
  number={6},
  pages={60008},
  year={2009}
}

@article{main2025distributed,
  title={Distributed quantum computing across an optical network link},
  author={Main, Dougal and Drmota, Peter and Nadlinger, David P and Ainley, Ellis M and Agrawal, Ayush and Nichol, Bethan C and Srinivas, Raghavendra and Araneda, Gabriel and Lucas, David M},
  journal={Nature},
  volume={638},
  number={8050},
  pages={383--388},
  year={2025},
  publisher={Nature Publishing Group UK London}
}

@inproceedings{boschero2025distributed,
  title={Distributed quantum computing: Applications and challenges},
  author={Boschero, Juan C and Neumann, Niels MP and van der Schoot, Ward and Sijpesteijn, Thom and Wezeman, Robert},
  booktitle={Intelligent Computing-Proceedings of the Computing Conference},
  pages={100--116},
  year={2025},
  organization={Springer}
}

@article{muralidharan2025simulation,
  title={The simulation of distributed quantum algorithms},
  author={Muralidharan, Sreraman},
  journal={The Journal of Supercomputing},
  volume={81},
  number={5},
  pages={645},
  year={2025},
  publisher={Springer}
}

@article{chefles2000quantum,
  title={Quantum state discrimination},
  author={Chefles, Anthony},
  journal={Contemporary Physics},
  volume={41},
  number={6},
  pages={401--424},
  year={2000},
  publisher={Taylor \& Francis}
}

@article{barnett2009quantum,
  title={Quantum state discrimination},
  author={Barnett, Stephen M and Croke, Sarah},
  journal={Advances in Optics and Photonics},
  volume={1},
  number={2},
  pages={238--278},
  year={2009},
  publisher={Optical Society of America}
}

@article{bergou2010discrimination,
  title={Discrimination of quantum states},
  author={Bergou, J{\'a}nos A},
  journal={Journal of Modern Optics},
  volume={57},
  number={3},
  pages={160--180},
  year={2010},
  publisher={Taylor \& Francis}
}

@article{bae2015quantum,
  title={Quantum state discrimination and its applications},
  author={Bae, Joonwoo and Kwek, Leong-Chuan},
  journal={Journal of Physics A: Mathematical and Theoretical},
  volume={48},
  number={8},
  pages={083001},
  year={2015},
  publisher={IOP Publishing}
}

@article{kimble2008quantum,
  title={The quantum internet},
  author={Kimble, H Jeff},
  journal={Nature},
  volume={453},
  number={7198},
  pages={1023--1030},
  year={2008},
  publisher={Nature Publishing Group}
}

@article{wehner2018quantum,
  title={Quantum internet: A vision for the road ahead},
  author={Wehner, Stephanie and Elkouss, David and Hanson, Ronald},
  journal={Science},
  volume={362},
  number={6412},
  pages={eaam9288},
  year={2018},
  publisher={American Association for the Advancement of Science}
}

@article{gupta2024bell,
  author  = {Gupta, Manu and Batin, Abdul Q. and Rajiuddin, {\relax Sk}  and Panigrahi, Prasanta K.},
  title   = {Bell State Discrimination and Error Correction},
  journal = {Asian Journal of Physics},
  volume  = {33},
  number  = {11},
  pages   = {683--688},
  year    = {2024},
}

@article{lim2025trade,
  title={Trade-off between information gain and disturbance in local discrimination of entangled quantum states},
  author={Lim, Youngrong and Hhan, Minki and Kwon, Hyukjoon},
  journal={Quantum Science and Technology},
  volume={10},
  number={2},
  pages={025048},
  year={2025},
  publisher={IOP Publishing}
}

@article{samal2010non,
  title={Non-destructive discrimination of Bell states by NMR using a single ancilla qubit},
  author={Samal, Jharana Rani and Gupta, Manu and Panigrahi, Prasanta K. and Kumar, Anil},
  journal={Journal of Physics B: Atomic, Molecular and Optical Physics},
  volume={43},
  number={9},
  pages={095508},
  year={2010},
  publisher={IOP Publishing}
}

@book{gottesman1997stabilizer,
  author={Gottesman, Daniel},
  title={Stabilizer codes and quantum error correction},
  publisher={California Institute of Technology},
  address = {Pasadena, CA, USA},
  year={1997},
}

@incollection{greenberger1989going,
  author    = {Greenberger, Daniel M. and Horne, Michael A. and Zeilinger, Anton},
  title     = {Going Beyond {B}ell's Theorem},
  booktitle = {Bell's Theorem, Quantum Theory and Conceptions of the Universe},
  pages     = {69--72},
  publisher = {Springer},
  address   = {Dordrecht, The Netherlands},
  year      = {1989}
}

@article{greenberger1990bell,
  title={Bell’s theorem without inequalities},
  author={Greenberger, Daniel M and Horne, Michael A and Shimony, Abner and Zeilinger, Anton},
  journal={American Journal of Physics},
  volume={58},
  number={12},
  pages={1131--1143},
  year={1990},
  publisher={American Association of Physics Teachers}
}

@article{dur2000three,
  title={Three qubits can be entangled in two inequivalent ways},
  author={D{\"u}r, Wolfgang and Vidal, Guifre and Cirac, J Ignacio},
  journal={Physical Review A},
  volume={62},
  number={6},
  pages={062314},
  year={2000},
  publisher={APS}
}

@article{walgate2000local,
  title={Local distinguishability of multipartite orthogonal quantum states},
  author={Walgate, Jonathan and Short, Anthony J and Hardy, Lucien and Vedral, Vlatko},
  journal={Physical Review Letters},
  volume={85},
  number={23},
  pages={4972},
  year={2000},
  publisher={APS}
}

@article{ghosh2001distinguishability,
  title={Distinguishability of Bell states},
  author={Ghosh, Sibasish and Kar, Guruprasad and Roy, Anirban and Sen, Aditi and Sen, Ujjwal and others},
  journal={Physical review letters},
  volume={87},
  number={27},
  pages={277902},
  year={2001},
  publisher={APS}
}

@article{horodecki2009quantum,
  title={Quantum entanglement},
  author={Horodecki, Ryszard and Horodecki, Pawe{\l} and Horodecki, Micha{\l} and Horodecki, Karol},
  journal={Reviews of modern physics},
  volume={81},
  number={2},
  pages={865--942},
  year={2009},
  publisher={APS}
}

@article{yin2017satellite,
  title={Satellite-based entanglement distribution over 1200 kilometers},
  author={Yin, Juan and Cao, Yuan and Li, Yu-Huai and Liao, Sheng-Kai and Zhang, Liang and Ren, Ji-Gang and Cai, Wen-Qi and Liu, Wei-Yue and Li, Bo and Dai, Hui and others},
  journal={Science},
  volume={356},
  number={6343},
  pages={1140--1144},
  year={2017},
  publisher={American Association for the Advancement of Science}
}

@article{welte2021nondestructive,
  title={A nondestructive Bell-state measurement on two distant atomic qubits},
  author={Welte, Stephan and Thomas, Philip and Hartung, Lukas and Daiss, Severin and Langenfeld, Stefan and Morin, Olivier and Rempe, Gerhard and Distante, Emanuele},
  journal={Nature Photonics},
  volume={15},
  number={7},
  pages={504--509},
  year={2021},
  publisher={Nature Publishing Group UK London}
}

@article{paige2020quantum,
  title={Quantum delocalized interactions},
  author={Paige, AJ and Kwon, Hyukjoon and Simsek, Selwyn and Self, Chris N and Gray, Johnnie and Kim, MS},
  journal={Physical Review Letters},
  volume={125},
  number={24},
  pages={240406},
  year={2020},
  publisher={APS}
}

@article{wootters2009no,
  title={The no-cloning theorem},
  author={Wootters, William K and Zurek, Wojciech H},
  journal={Physics Today},
  volume={62},
  number={2},
  pages={76--77},
  year={2009},
  publisher={AIP Publishing}
}

@article{javadi2024quantum,
  title={Quantum computing with Qiskit},
  author={Javadi-Abhari, Ali and Treinish, Matthew and Krsulich, Kevin and Wood, Christopher J and Lishman, Jake and Gacon, Julien and Martiel, Simon and Nation, Paul D and Bishop, Lev S and Cross, Andrew W and others},
  journal={arXiv preprint arXiv:2405.08810},
  year={2024}
}

@article{rossi2015entangled,
  title={Entangled quantum probes for dynamical environmental noise},
  author={Rossi, Matteo AC and Paris, Matteo GA},
  journal={Physical Review A},
  volume={92},
  number={1},
  pages={010302},
  year={2015},
  publisher={APS}
}

@article{carvalho2004decoherence,
  title={Decoherence and multipartite entanglement},
  author={Carvalho, Andr{\'e} RR and Mintert, Florian and Buchleitner, Andreas},
  journal={Physical review letters},
  volume={93},
  number={23},
  pages={230501},
  year={2004},
  publisher={APS}
}

@article{degen2017quantum,
  title={Quantum sensing},
  author={Degen, Christian L and Reinhard, Friedemann and Cappellaro, Paola},
  journal={Reviews of modern physics},
  volume={89},
  number={3},
  pages={035002},
  year={2017},
  publisher={APS}
}

@article{nolan2017quantum,
  title={Quantum Fisher information as a predictor of decoherence in the preparation of spin-cat states for quantum metrology},
  author={Nolan, Samuel P and Haine, Simon A},
  journal={Physical Review A},
  volume={95},
  number={4},
  pages={043642},
  year={2017},
  publisher={APS}
}

@article{huelga1997improvement,
  title={Improvement of frequency standards with quantum entanglement},
  author={Huelga, Susanna F and Macchiavello, Chiara and Pellizzari, Thomas and Ekert, Artur K and Plenio, Martin B and Cirac, J Ignacio},
  journal={Physical Review Letters},
  volume={79},
  number={20},
  pages={3865},
  year={1997},
  publisher={APS}
}

\end{document}